\documentclass[conference]{IEEEtran}
\usepackage{subfigure}
\usepackage{listings}
\usepackage{xcolor}
\usepackage{amsmath}
\usepackage{pifont}
\usepackage{graphicx}
	
\usepackage{soul}
\usepackage[shortlabels]{enumitem}

\usepackage{listings}
\usepackage{color}
\usepackage{tikz}
\usepackage{balance}
\usepackage{multirow}
\usepackage{comment}
\usepackage{amsmath}
\usepackage{graphicx}
\usepackage{wrapfig,lipsum,booktabs}
\usepackage{titlesec}
\usepackage[frozencache,cachedir=.]{minted}
\usepackage{dblfloatfix}

\newcommand{\cmark}{\ding{52}}
\newcommand{\xmark}{\ding{53}}

\definecolor{codegreen}{rgb}{0,0.6,0}
\definecolor{codegray}{rgb}{0.5,0.5,0.5}
\definecolor{codepurple}{rgb}{0.58,0,0.82}
\definecolor{backcolour}{rgb}{0.95,0.95,0.92}

\lstdefinestyle{mystyle}{
  backgroundcolor=\color{backcolour},   commentstyle=\color{codegreen},
  keywordstyle=\color{magenta},
  numberstyle=\tiny\color{codegray},
  stringstyle=\color{codepurple},
  basicstyle=\ttfamily\footnotesize,
  breakatwhitespace=false,         
  breaklines=true,                 
  captionpos=b,                    
  keepspaces=true,                 
  numbers=left,                    
  numbersep=5pt,                  
  showspaces=false,                
  showstringspaces=false,
  showtabs=false,                  
  tabsize=2,
}

\AtBeginDocument{%
  \providecommand\BibTeX{{%
    \normalfont B\kern-0.5em{\scshape i\kern-0.25em b}\kern-0.8em\TeX}}}

\IEEEoverridecommandlockouts
\begin{document}

\title{MCR-DL: Mix-and-Match Communication Runtime for Deep Learning
\thanks{This research is supported in part by NSF grants \#1818253, \#1854828, \#1931537, \#2007991, \#2018627, \#2112606, and XRAC grant \#NCR-130002.}}

\author{\IEEEauthorblockN{Quentin Anthony}
\IEEEauthorblockA{\textit{The Ohio State University}\\
Columbus, OH \\
anthony.301@osu.edu}
\and
\IEEEauthorblockN{Ammar Ahmad Awan}
\IEEEauthorblockA{\textit{Microsoft Corporation}\\
Redmond, WA \\
ammar.awan@microsoft.com}
\and
\IEEEauthorblockN{Jeff Rasley}
\IEEEauthorblockA{\textit{Microsoft Corporation}\\
Redmond, WA \\
jeff.rasley@microsoft.com}
\and
\IEEEauthorblockN{Yuxiong He}
\IEEEauthorblockA{\textit{Microsoft Corporation}\\
Redmond, WA \\
yuxhe@microsoft.com}
\and
\IEEEauthorblockN{Aamir Shafi}
\IEEEauthorblockA{\textit{The Ohio State University}\\
Columbus, OH \\
shafi.16@osu.edu}
\and
\IEEEauthorblockN{Mustafa Abduljabbar}
\IEEEauthorblockA{\textit{The Ohio State University}\\
Columbus, OH \\
abduljabbar.1@osu.edu}
\and
\IEEEauthorblockN{Hari Subramoni}
\IEEEauthorblockA{\textit{The Ohio State University}\\
Columbus, OH \\
subramoni.1@osu.edu}
\and
\IEEEauthorblockN{Dhabaleswar Panda}
\IEEEauthorblockA{\textit{The Ohio State University}\\
Columbus, OH \\
panda.2@osu.edu}
}

\maketitle

\setcounter{page}{1}

\begin{abstract}
In recent years, the training requirements of many state-of-the-art Deep Learning (DL) models have scaled beyond the compute and memory capabilities of a single processor, and necessitated distribution among processors. Training such massive models necessitates advanced parallelism strategies \cite{zero-offload, megatron-lm} to maintain efficiency. However, such distributed DL parallelism strategies require a varied mixture of collective and point-to-point communication operations across a broad range of message sizes and scales. Examples of models using advanced parallelism strategies include Deep Learning Recommendation Models (DLRM) \cite{dlrm} and Mixture-of-Experts (MoE) \cite{ds-moe-orig, ds-moe-latest}. Communication libraries' performance varies wildly across different communication operations, scales, and message sizes. We propose MCR-DL: an extensible DL communication framework that supports all point-to-point and collective operations while enabling users to dynamically mix-and-match communication backends for a given operation without deadlocks. MCR-DL also comes packaged with a tuning suite for dynamically selecting the best communication backend for a given input tensor. We select DeepSpeed-MoE and DLRM as candidate DL models and demonstrate a 31\% improvement in DS-MoE throughput on 256 V100 GPUs on the Lassen HPC system. Further, we achieve a 20\%  throughput improvement in a dense Megatron-DeepSpeed model and a 25\% throughput improvement in DLRM on 32 A100 GPUs with the Theta-GPU HPC system.
\end{abstract}


\section{Introduction}
\label{sec:intro}


Distributed DL has become the standard training method for many state-of-the-art vision \cite{gems}, language \cite{megatron-turing-nlg, meta-opt}, and recommendation \cite{dlrm-scale} DL models. As the largest models grow from hundreds of millions \cite{ResNet} to hundreds of billions of parameters \cite{megatron-turing-nlg}, new parallelization schemes have arisen to efficiently train DL models across thousands of processors \cite{zero, zero-offload, deepspeed}. While previous data-parallel DL models could heavily rely on a few collective operations (namely Allreduce), the model-parallel schemes of new models (e.g. sharding, pipeline and model parallelism, tensor slicing, etc) require a mixture of different collective and point-to-point operations \cite{dlrm, gshard, zero}. These advanced parallelization schemes rely heavily upon communication backends such as the NVIDIA Collectives Communication Library NCCL \cite{nccl} and CUDA-Aware MPI libraries \cite{openmpi, MVAPICH2}. However, modern communication backends have wildly varied performance characteristics across operations, within operations, and across releases (See Section \ref{sec:motivation} for a concrete example).


\subsection{Problem Statement}
\label{sec:problem-statement}

There are two primary drawbacks to existing distributed DL frameworks' communication: a lack of completeness in support for all communication operations/backends, and a lack of support for mixed-backend communication. Since modern distributed DL frameworks such as Horovod and PyTorch's Distributed module do not support all MPI or NCCL operations (e.g. vectored collectives such as Gatherv), DL researchers are required to either: \textbf{(Option 1):} implement their desired collectives via Point-to-Point operations (if point-to-point operations are supported in the chosen framework), or \textbf{(Option 2):} transfer tensors between the distributed DL framework and an external MPI Python wrapper such as mpi4py \cite{mpi4py}. Option 1 sacrifices the performance enhancements present in NCCL and most CUDA-Aware libraries, while option 2 introduces significant program complexity. For the second drawback, a lack of mixed-backend communication forces the user to decide where to sacrifice performance, since no communication backend performs all operations optimally (see Section \ref{sec:motivation} for a concrete example). These drawbacks bottleneck programmer productivity (e.g. a DL scientist must first implement an \textit{MPI\_Igather} before the intended optimization) and performance (e.g. NCCL performs well for Allreduce and MPI performs well for Alltoall. Which backend does one choose?), respectively.

\begin{figure*}[htbp]
  \begin{center}
      \mbox {
          \hspace{-1\columnsep}
          \subfigure[Proportion of computation to communication for distributed DL training]
          {
              \includegraphics[width=.4\textwidth,trim=2 2 2 2,clip]{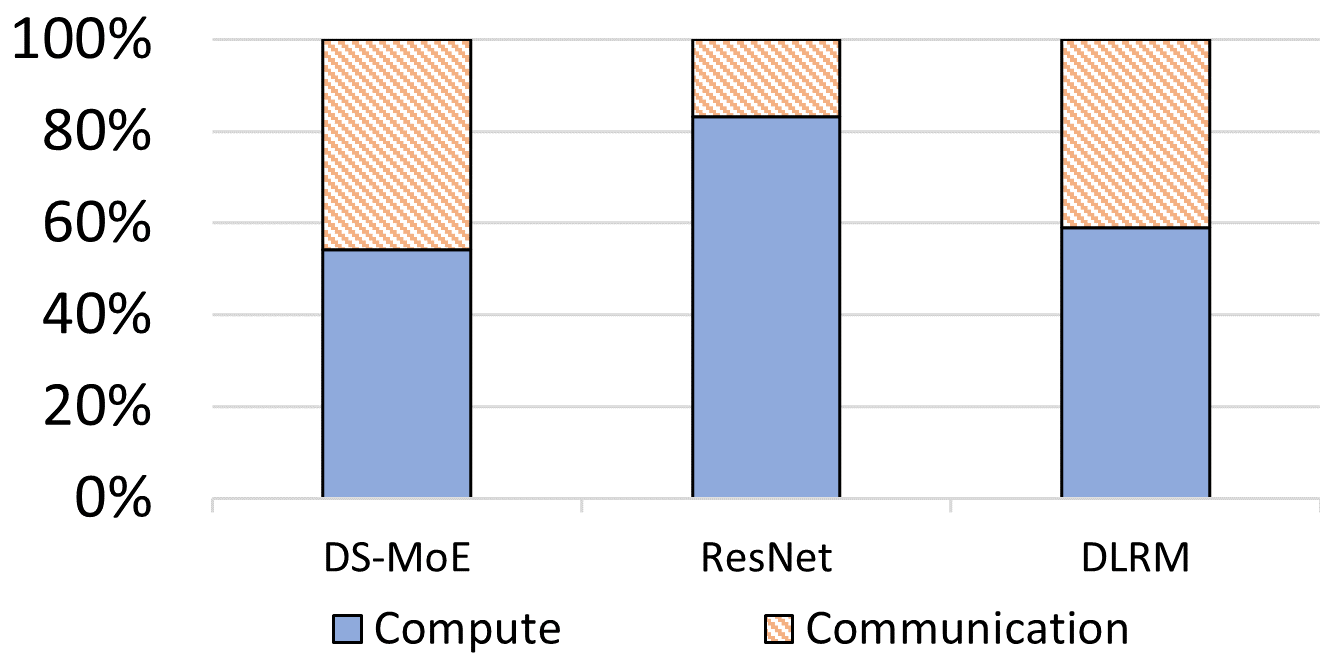}
              \label{fig:computevcomms}
          }
          \hspace{4ex}
          \subfigure[Breakdown of individual communication operations for distributed DL training]
          {
              \includegraphics[width=.4\textwidth,trim=2 2 2 2,clip]{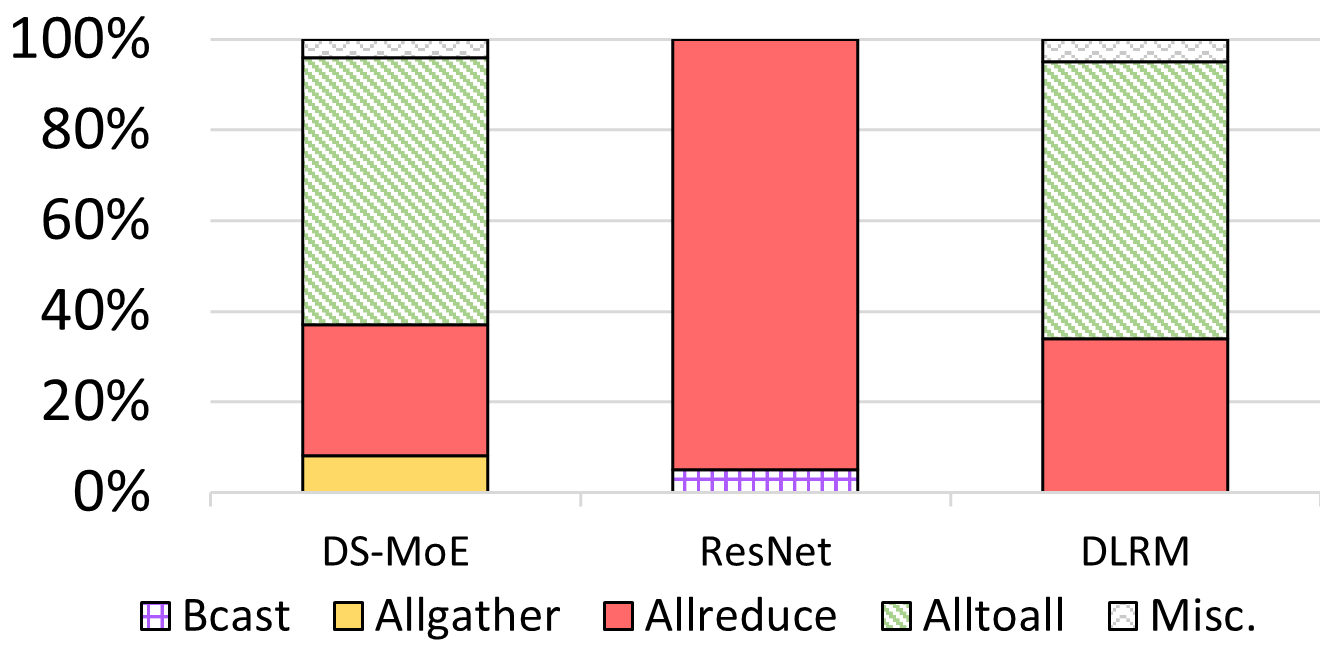}
              \label{fig:comms-breakdown}
          }
      }
      \vspace*{-0.5\baselineskip}
      \caption{Computation vs. Communication and breakdown of Communication operations breakdown for ResNet-50 (64 V100 GPUs on Lassen), DS-MoE (64 V100 GPUs on Lassen), and DLRM (32 A100 GPUs on Theta-GPU)}
      \label{fig:comms-profiles}
  \end{center}
\vspace*{-1\baselineskip}
\end{figure*}

\subsection{Proposed Solution}
\label{sec:proposed-solution}

We believe that a single unified interface between a given DL framework and the desired communication backend(s) (MPI, NCCL, etc) will alleviate these performance and productivity bottlenecks, while introducing the possibility of mixed backend communication (e.g. MPI Alltoall and NCCL Allreduce).

\begin{figure*}[hbtp]
  \begin{center}
      \mbox {
          \hspace{-1\columnsep}
          \subfigure[64 GPUs (16 node 4 ppn) - iAllreduce]
          {
              \includegraphics[width=0.4\linewidth,trim=2 2 2 2,clip]{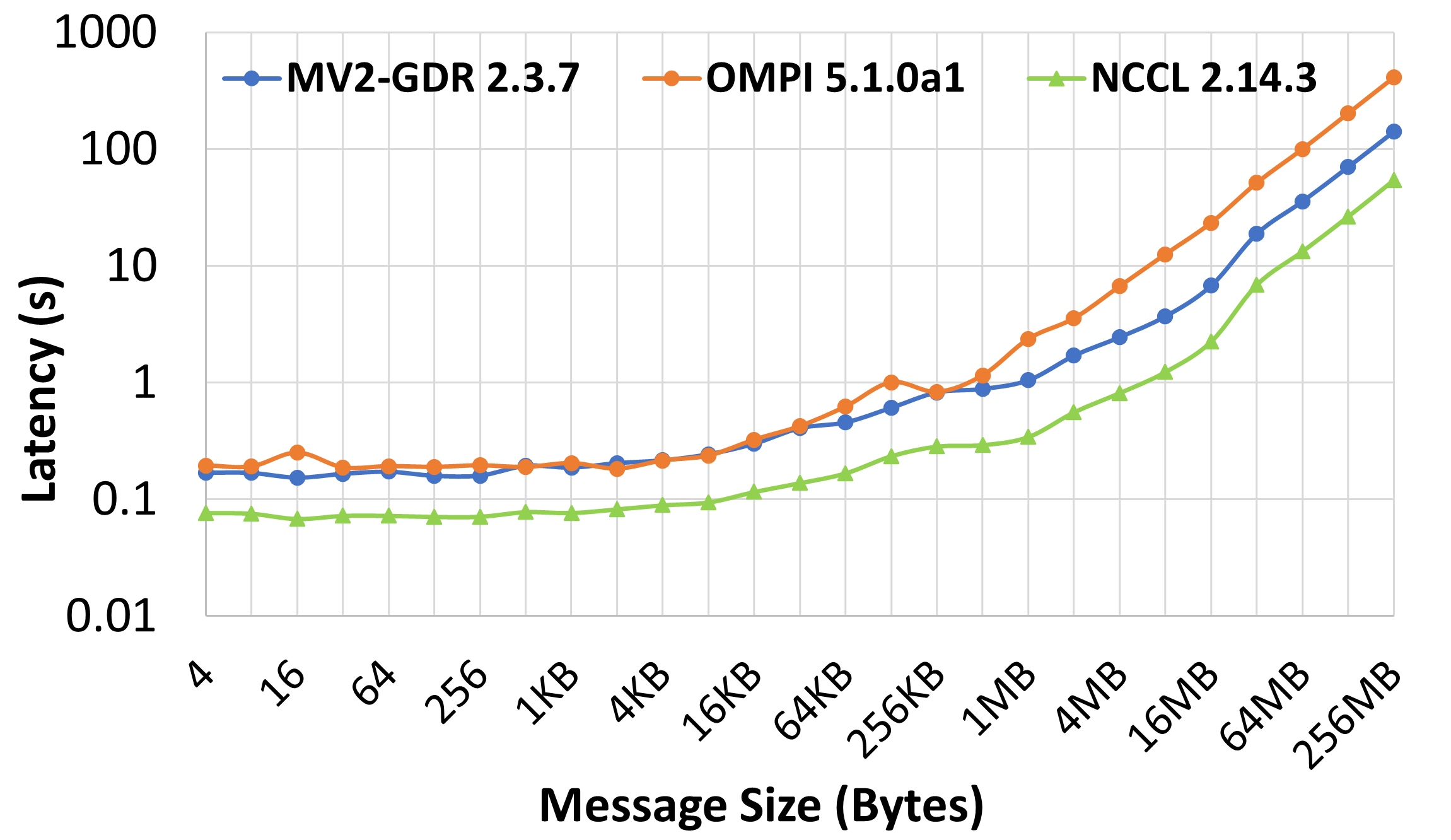}
              \label{fig:iallreduce-lassen-64gpu}
          }
          \hspace{4ex}
          \subfigure[64 GPUs (16 nodes 4 ppn) - Alltoall]
          {
              \includegraphics[width=0.4\linewidth,trim=2 2 2 2,clip]{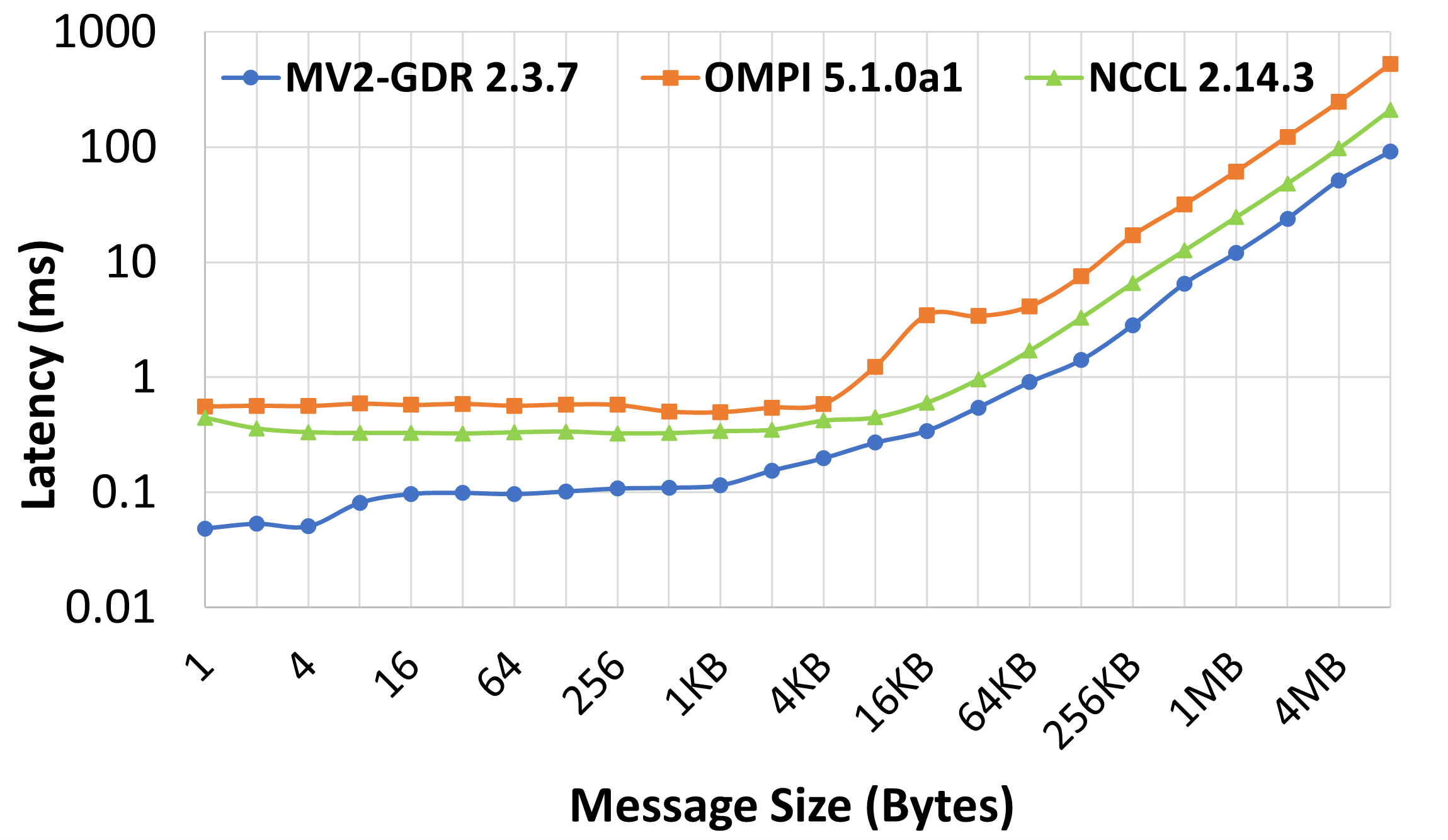}
              \label{fig:alltoall-lassen-64gpu}
          }
      }
      \vspace*{-0.5\baselineskip}
      \caption{Comparison of communication backends' collective performance on basic micro-benchmark with 64 V100 GPUs on Lassen}
      \label{fig:colls-lassen-64gpu}
  \end{center}
\vspace*{-1\baselineskip}
\end{figure*}

\renewcommand{\arraystretch}{1.5}
\begin{table*}[htb!]
\label{tab:related}
\resizebox{\textwidth}{!}{%
\begin{tabular}{|c|c|c|c|c|c|c|}
\hline
\multirow{2}{*}{\textbf{Studies}}                                                                  & \multicolumn{6}{c|}{\textbf{Features}}                                                                                                                                                                                                                                                                                                                                                                                                                                                                                                                                                                                                                                                                                                                                                                                                   \\ \cline{2-7} 
                                                                             & \textbf{\begin{tabular}[c]{@{}c@{}}Point-to-Point\end{tabular}} &
                                                        \textbf{\begin{tabular}[c]{@{}c@{}}Collectives\end{tabular}} &
                                                        \textbf{\begin{tabular}[c]{@{}c@{}}Vector Collectives\end{tabular}} &                             \textbf{\begin{tabular}[c]{@{}c@{}}Non-Blocking Operations\end{tabular}} &                              \textbf{\begin{tabular}[c]{@{}c@{}}Mixed-Backend Communication\end{tabular}} &  \textbf{\begin{tabular}[c]{@{}c@{}}Backend as a Class\end{tabular}} \\ \hline
\textbf{ \begin{tabular}[c]{@{}c@{}}Horovod\end{tabular}}                                                                                                &  \xmark                    &  \cmark   &  \xmark   &  NCCL Only                                                                      &  Experimental                                                                                    &  \xmark                                                                                      \\ \hline
\textbf{\begin{tabular}[c]{@{}c@{}}PyTorch Distributed Module\end{tabular}}     &  \cmark    & \cmark  &  \xmark                                                             & NCCL Only                                        & \xmark                          & \cmark                                                                                                                                                       \\ \hline
\textbf{LBANN}                       &  \cmark      &  \cmark  &  \xmark        &  \cmark                                                                                                     &  \xmark                                                                               &  \xmark                                                                                      \\ \hline
\textbf{mpi4py\cite{mpi4py}}              &  \cmark  &  \cmark  &  \cmark        & \cmark                                                                                                     & \xmark                                                                                          & \xmark                                                                                 \\ \hline
\textbf{\color[HTML]{009901}Proposed MCR-DL}         & \color[HTML]{009901}\cmark  & \color[HTML]{009901}\cmark         & \color[HTML]{009901}\cmark                                                                                                     & \color[HTML]{009901}\cmark                          & \color[HTML]{009901}\cmark                                             & \color[HTML]{009901}\cmark                                                                              \\ \hline
\end{tabular}
}
\vspace{1ex}
\caption{Features offered by MCR-DL compared to existing frameworks}
\vspace{-5ex}
\end{table*}

In this paper, we introduce and evaluate a \textbf{Mix-and-Match Communication Runtime for Deep Learning (MCR-DL)}. Specifically, MCR-DL is a lightweight unified interface between the DL framework (PyTorch) and any combination of ABI-compatible\footnote{An Application Binary Interface (ABI), is the low-level interface between two program modules. An ABI determines such details as how functions are called and the size, layout, and alignment of datatypes. With ABI-compatibility, programs conform to the same set of runtime conventions.} communication backends. MCR-DL users can dynamically switch between communication backends during distributed DL training. MCR-DL supports many existing communication backends (by implementing them as a high-level backend class), and provides an extensible design to enable new communication backends and performance optimizations.

\subsection{Motivation}
\label{sec:motivation}

First, we profiled the computation and communication overhead of three representative DL models: DLRM and DeepSpeed-MoE (state-of-the-art hybrid-parallel DL models), and ResNet-50 (established data-parallel DL model). The overall computation vs. communication split as well as communication breakdown profiles are depicted in Figure \ref{fig:comms-profiles}. First, we note that data-parallelism is strongly compute-dominated, and its communication overhead is almost entirely made up of Allreduce. Therefore, data-parallel applications like ResNet-50 are able to achieve the best performance on existing monolithic distributed DL frameworks, and the choice of communication backend is simply determined by whichever library has the fastest CUDA-Aware Allreduce. We note that MCR-DL is still applicable to data-parallel frameworks with tuning (See Section \ref{sec:design-tuning} and Table \ref{tab:tuning-table} for details), but due to their much lower communication overhead, the benefits are marginal.

However, DLRM and DS-MoE have a significantly higher communication overhead at scale. Further, their communication operation requirements are heterogeneous. Therefore, there is a lot of room for mixing backends according to their strengths in order to improve training throughput.

\textit{Consider the case of DS-MoE. Given the communication breakdown in Figure \ref{fig:comms-breakdown} and the collective performance in Figure \ref{fig:colls-lassen-64gpu}, which communication backend should be used?} A myriad of application questions would need to be answered such as which collectives DS-MoE uses, their relative frequencies, and the range of message sizes for each collective. Any decision on a single communication backend will lose out on some collectives and at some message ranges. Specifically, since DS-MoE relies mostly on Allreduce and Alltoall, we could refer to Figure \ref{fig:colls-lassen-64gpu} and reduce communication overhead by applying MVAPICH2-GDR for Alltoall and NCCL for Allreduce. However, such a decision will need to be reevaluated at each subsequent release cycle of the communication backends. If the user is able to dynamically switch among communication backends, they could squeeze more performance out of their application while reducing the setup cost of changing communication backends if future communication backend releases change.

\subsection{Contributions}



Our contributions are as follows:

\begin{enumerate}

    \item[\textbf{C1)}] We proposed, designed, and evaluated MCR-DL: an extensible, scalable API for DL communication operations. MCR-DL supports all point-to-point and collective communication operations on PyTorch tensors, and all collective communication libraries (Section \ref{sec:design-api})
    
    \item[\textbf{C2)}] We enabled deadlock-free mixed-backend DL communication via fine-grained synchronization techniques (Section \ref{sec:design-mixed-backend})
    
    \item[\textbf{C3)}] We fully implemented MCR-DL in a C++ backbone underneath a thin Python layer, and achieved a maximum of 5\% overhead (compared to a pure micro-benchmark written in C) for small messages and a 1\% overhead for large messages (down from 18\% and 4\% in PyTorch distributed, respectively) (Figure \ref{fig:overhead})
    
    \item[\textbf{C4)}] MCR-DL offers up to 31\% throughput improvement (12\% in scaling efficiency) in DeepSpeed-MoE and 25\% throughput improvement (14\% improvement in scaling efficiency) in DLRM by dynamically selecting the best-performing communication backend at each scale and message size (Figures \ref{fig:moe-results} and \ref{fig:dlrm-results})
    
    \item[\textbf{C5)}] Define and implement a tuning framework within MCR-DL that enables the best communication backend to be automatically selected for each communication operation (Section \ref{sec:design-tuning})
    
    \item[\textbf{C6)}] Demonstrate the extensibility of MCR-DL by adding support for communication compression, logging, and tensor fusion (Section \ref{sec:design-extensibility})
    
    
\end{enumerate}
\section{Related Work}
\label{sec:related}

\subsection{DL Communication Framework Design}

DeepSpeed \cite{deepspeed} uses PyTorch's distributed module \cite{pytorch-dist} to implement optimized DL communication at extreme scales. Recently, DeepSpeed has added support for Mixture-of-Experts (MoE) DL models \cite{ds-moe-latest, ds-moe-orig}. Horovod \cite{horovod} is a data-parallel focused framework that experimentally supports mixed communications without deadlock-avoidance support. The Livermore Big Artificial Neural Network Toolkit (LBANN) is an HPC-centric distributed DL framework that supports multiple parallelism levels. The MPI for Python package \cite{mpi4py} supplies Python bindings for the MPI standard. Our work competes with these works by seeking to unify communication calls into a single interface built atop PyTorch.

\subsection{Mixing MPI with an External Framework}

The work in \cite{jose-upc} combined an MPI runtime with UPC in a deadlock-free architecture by unifying the runtimes. The resulting runtime shared resources between MPI and UPC to avoid data-dependencies. In recent releases, the MVAPICH2-GDR \cite{MVAPICH2} CUDA-Aware MPI library has added support for NCCL collectives. However, this support is not optimized for non-blocking communication operations like those required by DLRM. Aluminum \cite{aluminum} is a DL-focused communication library built on MPI and NCCL, but is focused on latency-bound communication operations. Our work is complementary to the above works, since we choose the best backend for each communication operation.

\subsection{Scaling Mixture-of-Experts and DLRM Models}

The work in \cite{gshard} scaled a 600 billion parameter Mixture-of-Experts (MoE) model to 2048 TPU v3 processors. DeepSpeed has recently added support for MoE DL models \cite{ds-moe-orig} and scaled beyond a trillion parameters \cite{ds-moe-latest}. MoE models are gradually being applied to other domains such as vision \cite{vit-moe}. DLRM \cite{dlrm} has scaled beyond a trillion parameters with 4D parallelism techniques \cite{dlrm-scale}. We demonstrate that our work further improves the scaling behavior of these complex parallel DL models.
\section{Background}
\label{sec:background}
\subsection{DL Training}

Distributed DL can take several forms: data-parallelism, model-parallelism, and hybrid-parallelism. Data-parallism places a full model replica on each processor, and splits the training data among processors. Model parallelism splits the model across processors, and propagates each data sample through each device. Hybrid-parallelism splits the model across sets of processors, and splits the training data among complete-model sets of processors. There are tradeoffs for each parallelism scheme: data-parallelism is the simplest and has low communication overhead but is restricted to models that fit in processor memory. Hybrid and model-parallelism can accommodate any model size, but can require complex communication with high overheads. All distributed DL schemes are increasingly deployed on HPC systems \cite{mtf, megatron-turing-nlg}.

\subsection{Distributed DL Frameworks}

Horovod is a distributed DL framework with a focus on distributed data-parallelism to train DNNs \cite{horovod}. As such, Horovod primarily relies on Allreduce and Bcast collectives. Due to Horovod's focus, they provide a simple API, quick installations, and powerful data-parallel optimizations and profiling tools. Horovod supports many major DL frameworks and communication backends, including MPI and NCCL \cite{nccl}. 

PyTorch's distributed module is a built-in communication API within the PyTorch \cite{pytorch} DL framework. PyTorch distributed supports most communication operations, and contains several optimizations for distributed training (e.g. mixed-precision, gradient bucketing, sharded optimizer states). While official PyTorch wheels come packaged with the NCCL backend, other backends require a PyTorch source installation. 

DeepSpeed is a distributed DL framework built atop PyTorch's distributed module. DeepSpeed's focus is on efficient training of large-scale models that don't fit into a single processor's memory. A myriad of parallelism schemes and optimizer sharding techniques are included in DeepSpeed. 

\subsection{Communication Backends}

MPI is a parallel programming standard that enables processes to communicate with each other. CUDA-aware MPI libraries such as SpectrumMPI ~\cite{spectrum-mpi}, OpenMPI ~\cite{openmpi}, and MVAPICH2 ~\cite{MVAPICH2} provide optimized support for heterogeneous systems containing GPUs. GPU communication optimizations such as staging, CUDA Inter-Process Communication (IPC), and GPUDirect RDMA enable MPI libraries to provide superior performance across different combinations of GPU and interconnect \cite{Kawthar:IWOPH19}.

NCCL implements optimized collective communication patterns for NVIDIA GPUs ~\cite{nccl}. The various collective communication primitives found in NCCL are: Allgather, Allreduce, Reduce, ReduceScatter, Alltoall, Point-to-Point, and Broadcast. NCCL is not MPI-compliant, however, and does not provide support for many common MPI operations such as gather, scatter, and variable message-size collectives. Microsoft's Synthesized Collective Communication Library (MSCCL) \cite{sccl} creates custom collective algorithms for a given hardware topology. MSCCL supports both AMD and NVIDIA GPUs, and supports all major collective operations. 

\subsection{Mixture-of-Experts}

Mixture-of-experts (MoE) is an ensemble machine learning technique where a collection of "expert" feed-forward networks (FFNs) are trained on subtasks of the problem. Only a few experts are applied to a given data sample. In recent years, the MoE technique has been applied to transformer DL models in an effort to increase the model size (and therefore accuracy) while lessening the computational burden. MoE models require less computation to train than equivalent standard (i.e. "dense") models because each token only propagates through an expert subset of the full model. Incoming tokens are routed to existing expert FFNs via a gating function, and this routing as well as its subsequent combination of FFN outputs require Alltoall operations. Such Alltoall operations scale with the number of devices, and quickly become a dominant communication overhead at large scales. The distributed DL frameworks DeepSpeed \cite{ds-moe-latest,ds-moe-orig} and Fairseq \cite{fairseq-moe} have recently added support for MoE transformer models.

\subsection{Deep Learning Recommendation Models}

Deep Learning Recommendation Models (DLRMs) are a family of recommendation models that rely upon at least one deep neural network (DNN) \cite{dlrm, dlrm-scale}. Such models are composed of sparse embedding tables and dense multilayer perceptrons (MLPs). Note that a MLP is a special case of an FFN where every layer is fully connected to the next layer in the network. While sparse categorical data must be processed via embedding lookups (and are memory-bound), dense continuous data is fed through the bottom MLPs (and are compute-bound). The MLPs are trained via data-parallelism, and hence depend on Allreduce. The embedding tables are split across processes, and must be shuffled with an Alltoall prior to being fed into the top MLP. Each batch's Alltoall operation is overlapped with the previous top MLP's forward pass from the previous batch, which necessitates non-blocking Alltoall. 
\section{Challenges}

The key challenge addressed in this paper is: \textit{Can we improve the interface between a DL framework and communication backends with a single unified framework built on top of PyTorch?}. We seek to create an extensible framework that encapsulates all MPI and NCCL functionality. To answer this broad question, we solve the following concrete challenges:

\begin{itemize}
    \item What are the key communication needs of modern distributed DL models and frameworks? Do existing distributed DL frameworks provide these needs?
    
    \item Can a unified framework improve rapid prototyping for DL parallelism schemes while enabling mixed-backend communications?
    
    \item What benefits can mixed-backend communications provide to improve DL training throughput?
\end{itemize}
\section{Design}
\label{sec:design}

MCR-DL is split into a C++ implementation layer underneath a thin Python wrapper. Each backend is implemented as an object of a class, and implements the MCR-DL API in accordance with each backend's requirements.

\subsection{MCR-DL API}
\label{sec:design-api}

MCR-DL implements all communication operations as depicted below in Listing \ref{lst:api}. 

\begin{lstlisting}[language=Python, linewidth=9cm, xleftmargin=2.0ex, label={lst:api}, caption=High-level MCR-DL API, otherkeywords={torch.Tensor}, deletendkeywords={input}, literate={get\_size}{\bfseries get\_size}{8}{get\_rank}{\bfseries get\_rank}{8}{wait}{\bfseries wait}{4}{init}{\bfseries init}{4}{get\_backends}{\bfseries get\_backends}{12}{finalize}{\bfseries finalize}{9}{synchronize}{\bfseries synchronize}{11}{send}{\bfseries send}{4}{recv}{\bfseries recv}{4}{all\_to\_all}{\bfseries all\_to\_all}{10}{all\_to\_allv}{\bfseries all\_to\_allv}{11}{all\_to\_all\_single}{\bfseries all\_to\_all\_single}{17}{all\_reduce}{\bfseries all\_reduce}{10}{all\_gather}{\bfseries all\_gather}{10}{gather}{\bfseries gather}{6}{scatter}{\bfseries scatter}{7}{reduce}{\bfseries reduce}{6}{reduce\_scatter}{\bfseries reduce\_scatter}{14}{gatherv}{\bfseries gatherv}{7}{scatterv}{\bfseries scatterv}{7}{bcast}{\bfseries bcast}{5}{all\_gatherv}{\bfseries all\_gatherv}{11}]
def get_backends()
def init(list<str> backends)
def finalize(list<str> backends)
def synchronize(list<str> backends)
def get_size(str backend)
def get_rank(str backend)
def send(str backend, torch.Tensor t, int rank, bool async_op)
def recv(str backend, torch.Tensor t, int rank, bool async_op)
def all_to_all_single(str backend, torch.Tensor output, torch.Tensor input, bool async_op)
def all_to_all(str backend, list<torch.Tensor> output, list<torch.Tensor> input, bool async_op)
def all_reduce(str backend, torch.Tensor output, ReduceOp op, bool async_op)
def all_gather(str backend, torch.Tensor output, torch.Tensor input, bool async_op)
def gather(str backend, torch.Tensor output, int root, bool async_op)
def scatter(str backend, torch.Tensor output, int root, bool async_op)
def reduce(str backend, torch.Tensor output, int root, ReduceOp op, bool async_op)
def reduce_scatter(str backend, torch.Tensor output, int root, ReduceOp op, bool async_op)
def bcast(str backend, torch.Tensor output, int root, bool async_op)
def gatherv(str backend, torch.Tensor output, int root, list<int> rcounts, list<int> displs, bool async_op)
def scatterv(str backend, torch.Tensor output, int root, list<int> scounts, list<int> displs, bool async_op)
def all_to_allv(str backend, torch.Tensor output, torch.Tensor input, list<int> scounts, list<int> rcounts, list<int> def sdispls, list<int> rdispls, bool async_op)
def all_gatherv(str backend, torch.Tensor output, int root, list<int> rcounts, list<int> displs, bool async_op)
\end{lstlisting}

\begin{lstlisting}[language=Python,label={lst:productivity}, linewidth=9cm, xleftmargin=2.0ex, deletendkeywords={input}, otherkeywords={torch.int64}, literate={mcr\_ll}{\bfseries mcr\_dll}{6}, caption=Example of simplified prototyping with MCR-DL]
# Before MCR-DL
def allgather_host(self,
                   comm,
                   cupy_sign,
                   cupy_rbuf_sign,
                   cupy_scale,
                   cupy_rbuf_scale):

    # 1. Convert cupy to numpy
    numpy_rbuf_sign = np.zeros(
        [comm.Get_size(),
         cupy_sign.size],
        dtype=cupy_sign.dtype)
    numpy_rbuf_scale = np.zeros([comm.Get_size(),
                                           1],
                                          dtype=cupy_scale.dtype)

    numpy_sign = cupy.asnumpy(cupy_sign)
    numpy_rbuf_sign = cupy.asnumpy(cupy_rbuf_sign)
    numpy_scale = cupy.asnumpy(cupy_scale)
    numpy_rbuf_scale = cupy.asnumpy(cupy_rbuf_scale)
    cupy.cuda.get_current_stream().synchronize()

    # 2. Communicate numpy buffers
    comm.Allgather(numpy_sign, numpy_rbuf_sign)
    comm.Allgather(numpy_scale, numpy_rbuf_scale)
    comm.Barrier()

    # 3. Convert numpy back to cupy
    cupy_sign = cupy.asarray(numpy_sign)
    cupy_rbuf_sign = cupy.asarray(numpy_rbuf_sign)
    cupy_scale = cupy.asarray(numpy_scale)
    cupy_rbuf_scale = cupy.asarray(numpy_rbuf_scale)
    cupy.cuda.get_current_stream().synchronize()

    return cupy_sign, cupy_rbuf_sign, cupy_scale, cupy_rbuf_scale
	
	
# After MCR-DL
def allgather_host(self,
                   comm,
                   sign,
                   rbuf_sign,
                   scale,
                   rbuf_scale):
             
	comm.all_gather_base(rbuf_sign, sign)
	comm.all_gather_base(rbuf_sign, sign)
	
	return sign, rbuf_sign, scale, rbuf_scale
\end{lstlisting}

There are a few key takeaways from this API listing:

\begin{itemize}
    \item All operations take either a single \textbf{backend} string that matches an underlying backend class (e.g. "mv2-gdr", "nccl", etc) or a special backend flag "auto", which will dynamically choose the best message size for a given scale and message size if tuning tables are available. (Note: MCR-DL comes packaged with a tuning suite which first runs communication operation benchmarks for each backend, and uses this data to map each message size, scale, and operation to a given backend. This optimal backend choice is then used at runtime if "auto" is chosen)
    \item We conform to the PyTorch distributed module API conventions when possible to ease code refactoring to MCR-DL. An example of this is \textit{all\_to\_all}, which shuffles lists of tensors rather than individual tensor elements. This is a common usecase in distributed PyTorch applications. Another example is \textit{all\_to\_all\_single}, which directly shuffles the tensor elements themselves on each rank.
    \item Vectored collectives (e.g. gatherv/scatterv) and non-blocking collectives are supported for all backends.
\end{itemize}

\subsection{Advanced Communication Support}
Most distributed DL frameworks do not support the full underlying communication backend, only the operations that matter for DL parallelism (e.g. Allreduce). If a user needs a communication operation that is not currently supported by their distributed DL framework (e.g. advanced parallelism or data processing), they would need to sacrifice performance or productivity as mentioned in Section \ref{sec:motivation}. 

\begin{figure}[htbp]
\centering
    \includegraphics[width=.8\linewidth]{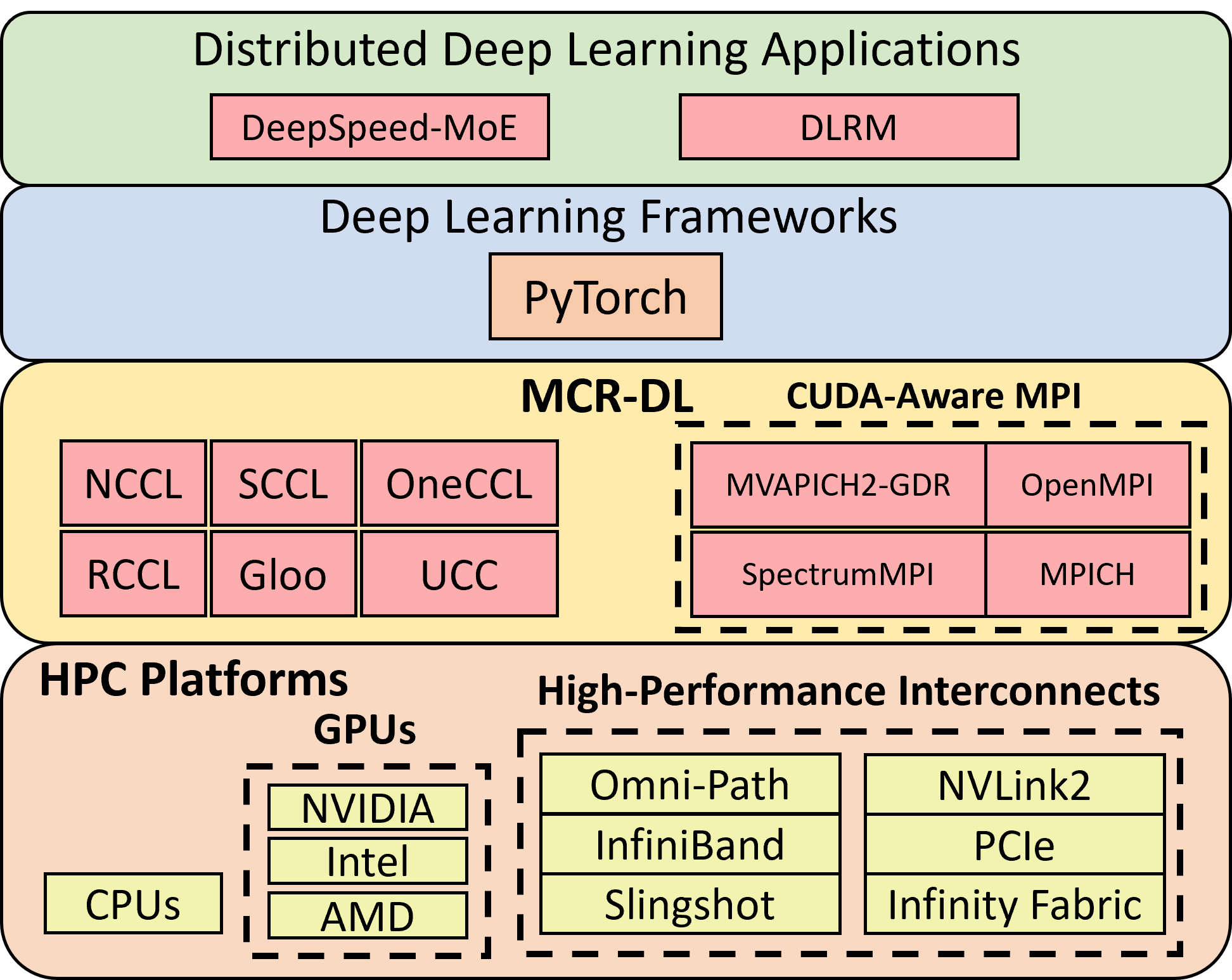}
    \caption{The MCR-DL Software Stack. MCR-DL is a thin layer between a target DL framework and the HPC system, and supports any number of stream-aware communication backends along with CUDA-Aware MPI.}
    \vspace{-3ex}
    \label{fig:dl-stack}
\end{figure}

MCR-DL is a thin layer atop each currently-supported backend, and fully implements each backend on PyTorch tensors (See Figure \ref{fig:dl-stack} for the software stack). The MCR-DL "Backend" class can be easily extended to new communication backends such as MSCCL \cite{sccl}, Gloo, oneAPI, etc. 

\subsection{Synchronization}
\label{sec:design-synchronization}

One of the most important design considerations for a distributed framework is synchronization. We seek to add enough synchronization to rid the programmer of having to frequently debug deadlocks and data validation issues, while achieving enough overlap to maintain high performance at scale. With the right synchronization strategy, we are able to efficiently both overlap computation with communication, and overlap across communication backends without deadlocks or data validation issues.

\begin{lstlisting}[language=Python,label={lst:synch}, linewidth=9cm, xleftmargin=2.0ex, deletendkeywords={input}, otherkeywords={torch.int64}, literate={mcr\_dl}{\bfseries mcr\_dl}{6}, caption=Example of available overlap between communication and computation in a DL setting]
import torch
import mcr_dl

def tensor():
    return torch.rand(1,1)

x = tensor().cuda()
y = tensor().cuda()
mcr_dl.init('nccl')

h = mcr_dl.all_reduce('nccl', x, async_op=True)
y = y + y
h.wait('nccl')
result = x + y
\end{lstlisting}
\vspace{-3ex}

\begin{figure}[htbp]
  \begin{center}
      \mbox {
          \hspace{-1\columnsep}
          \subfigure[Naive synchronization]
          {
            \includegraphics[width=.35\linewidth]{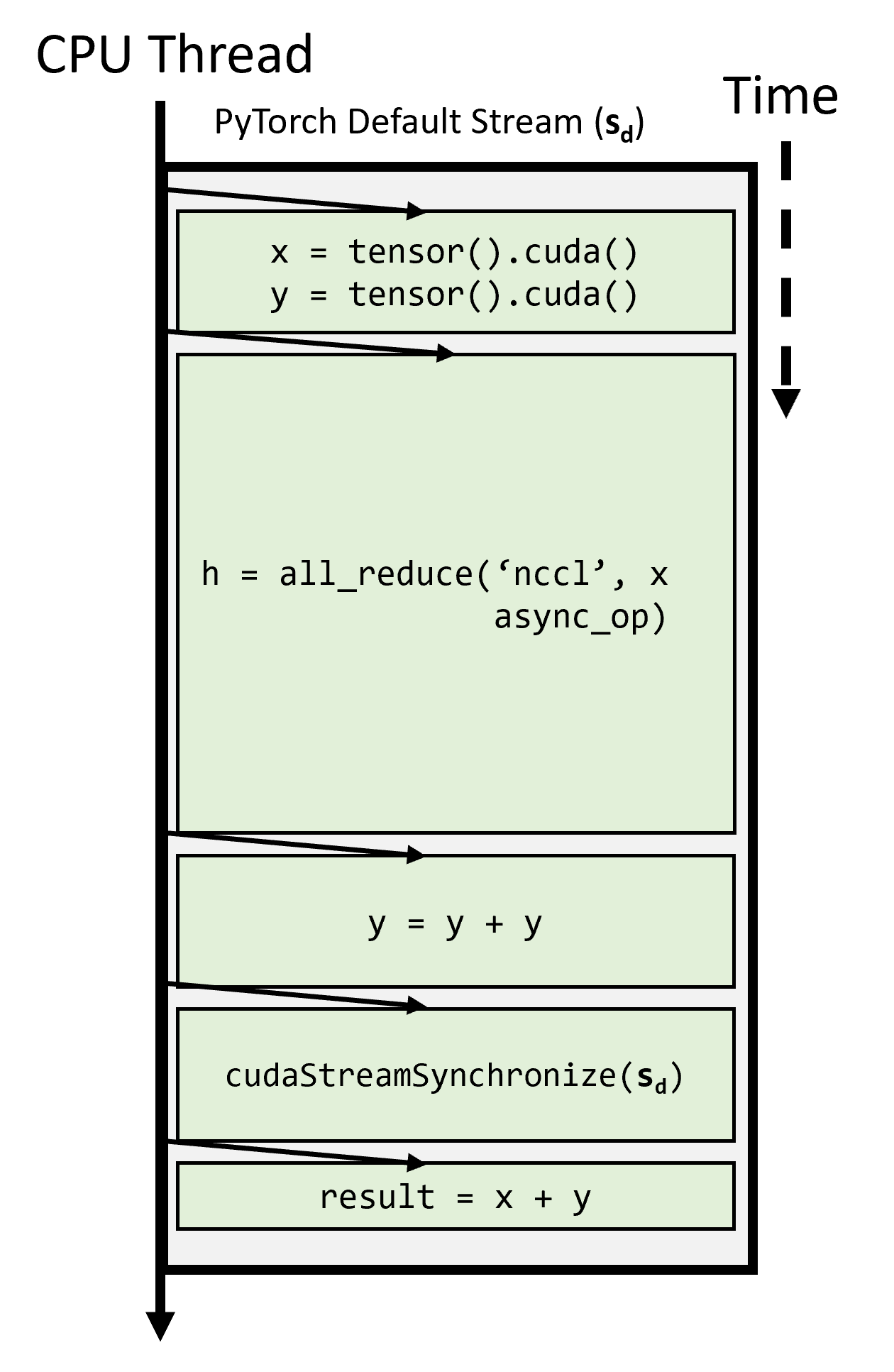}
            \label{fig:default-synch}
          }
          \subfigure[Synchronization in MCR-DL]
          {
            \includegraphics[width=.54\linewidth]{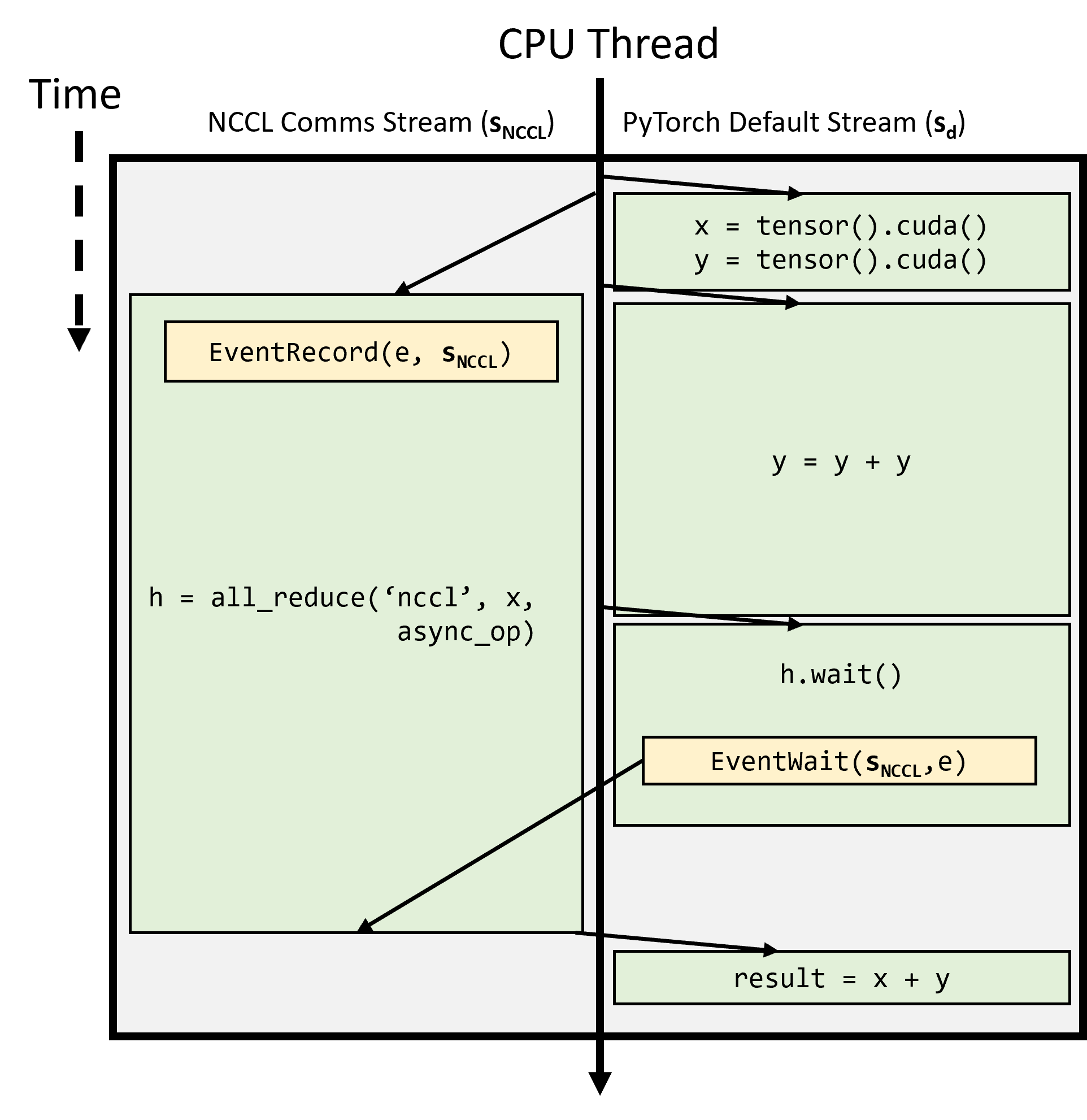}
            \label{fig:MCR-DL-synch}
          }
      }
      \vspace*{-0.5\baselineskip}
      \caption{Synchronization diagrams of Listing \ref{lst:synch} for the naive scheme and MCR-DL's fine-grained CUDA event scheme}
      \label{fig:single-backend-synchs}
  \end{center}
\vspace*{-1\baselineskip}
\end{figure}

First consider a naive synchronization scheme where a) all communication operations are posted to the PyTorch default stream, and b) we synchronize operations with cudaStreamSynchronize on that stream. We demonstrate the behavior of this scheme in Listing \ref{lst:synch}, a prototypical example of available communication/computation overlap faced in distributed DL. The resulting serial execution is depicted in Figure \ref{fig:default-synch}\footnote{The length of operation boxes in Figures is purely for synchronization discussion}. In MCR-DL, we exploit communication/computation overlap by creating a pool of \textbf{communication streams} for each backend. These streams are managed internally to MCR-DL. Communication operations posted to a backend's stream(s) are synchronized with fine-grained CUDA events. For figure \ref{fig:MCR-DL-synch}, this translates to: \textbf{(1):} An \textit{all\_reduce(x)} operation is posted to a NCCL communication stream in MCR-DL, and a distributed work handle is stored in \textit{h}, \textbf{(2):} MCR-DL records a CUDA event \textit{e} onto the communication stream and begins executing the \textit{all\_reduce(x)}, \textbf{(3):} the PyTorch default stream is able to progress with operations unrelated to \textit{x}, \textbf{(4):} when a data-dependency on \textit{x} is encountered, the user must call \textit{wait()} on the work handle \textit{h}, which MCR-DL uses internally to wait on the prior event \textit{e}. 

This scheme is similar to PyTorch's distributed module, but there are a few key implementation details that enable greater performance: \textbf{(1):} The use of multiple streams enables concurrent small-message operations (concurrent large-message operations are bandwidth-bound and show no benefit), \textbf{(2):} Instead of having an overall communication stream, each backend contains its own stream for overlap across backends. This synchronization behavior is extended to multiple backends in MCR-DL, which we will now discuss.

\subsection{Mixed-Backend Communications}
\label{sec:design-mixed-backend}

Since MCR-DL is a thin layer atop communication backends, we can pass the desired backend for any given communication operation dynamically within a Python script. An example of this is depicted below in Listing \ref{lst:mixed}.

\begin{lstlisting}[language=Python,label={lst:mixed}, linewidth=9cm, xleftmargin=2.0ex,  deletendkeywords={input}, otherkeywords={torch.int64}, literate={mcr\_dl}{\bfseries mcr\_dl}{6}, caption=Example of explicit mixed-backend communications in MCR-DL. All inter-backend synchronization is performed internally. MCR-DL can dynamically choose the best backend to use at runtime if \textit{'auto'} is passed as the backend (See Section \ref{sec:design-tuning})]
import torch
import mcr_dl

def tensor():
    return torch.rand(1,1)

x = tensor().cuda()
y = tensor().cuda()
z = tensor().cuda()
mcr_dl.init(['nccl', 'mpi'])

h1 = mcr_dl.all_reduce('nccl', x, async_op=True)
h2 = mcr_dl.all_reduce('mpi', y, async_op=True)
z = z + z
h1.wait()
h2.wait()
result = x + y + z
\end{lstlisting}

\begin{figure}[htbp]
   \centering
        \includegraphics[width=.89\linewidth]{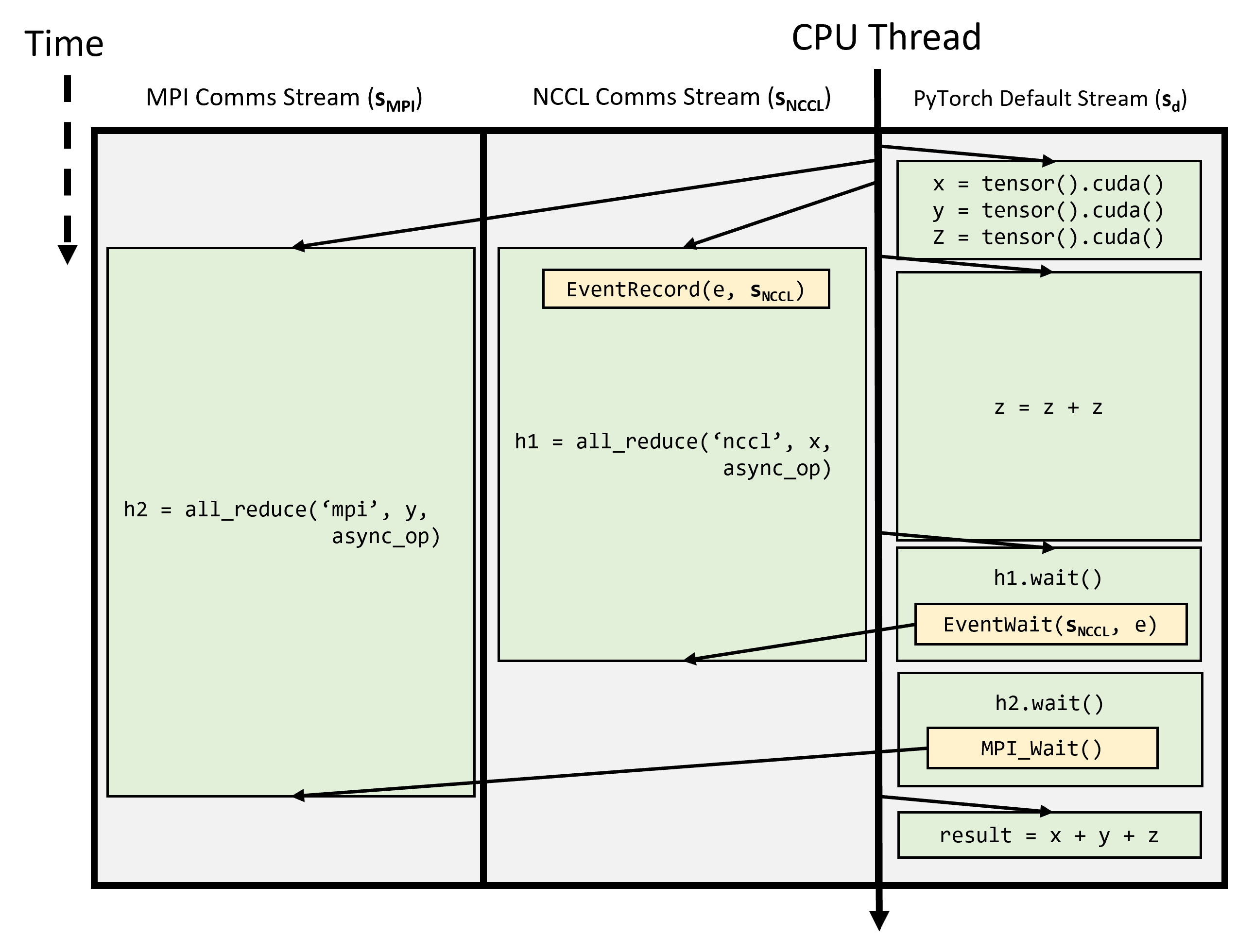}
        \caption{In MCR-DL, communication backends can be explicitly chosen, or users can dynamically choose the best backend for a given operation with "auto"} 
        \label{fig:MCR-DL multisynch}
        \vspace{-2ex}
\end{figure}

However, each communication backend conforms to its own synchronization scheme. NCCL and its derivatives are synchronized on the CUDA streams, while MPI is synchronized on a host thread. If we are to mix backends without deadlocks, we will need to loop over each implemented backend and synchronize with their respective thread/stream. For the mixture of CUDA-Aware MPI and non-blocking NCCL, for example, this entails a call to CUDA-event based synchronization for NCCL (as discussed in section \ref{sec:design-synchronization}), followed by an \textit{MPI\_Wait} for MPI. Handling CUDA-aware MPI is a challenge since CUDA streams are not exposed by MPI to the application, this leads to two options (which MCR-DL provide at the initialization of an MPI backend): \textbf{(1):} Allow MPI to handle all streams, which sacrifices some MCR-DL overlap across backends, but preserves multiple CUDA stream logic (if it exists) within MPI. \textbf{(2):} Intercept calls to cudaStreamCreate and manage streams in MCR-DL, which exploits overlap across backends, but could potentially lead to deadlocks if multi-stream logic is used in MPI\footnote{In our experiments, we find that the best choice for this option is dependent on the MPI library}. An example of streams managed by MCR-DL is depicted by Figure \ref{fig:MCR-DL multisynch}. For ease of synchronization, every work handle's \textit{wait()} call waits on the PyTorch default stream (i.e. synchronization purely between communication streams is not supported). We note that stream-aware MPI like the implemention by MPICH \cite{stream-mpich} allows MCR-DL to fully overlap communication backends by self-managing streams.

While Figure \ref{fig:MCR-DL multisynch} depicts the mixture of a stream-aware backend (NCCL) and a backend without streams exposed to the user (MPI), the combination of any number of stream-aware backends (NCCL, SCCL, etc) is supported in MCR-DL and synchronized with CUDA events. Further, the combination of ABI-compatible MPI backends is supported\footnote{In our experiments, we found that mixing at most one non-stream-aware backend is optimal for overlap}. In our experiments, the initialization overhead for multiple communication libraries is negligible after being amortized over a few $(<10)$ DL training steps.


\subsection{Communication Optimization Extensibility}
\label{sec:design-extensibility}

\begin{figure}[htbp]
   \centering
        \includegraphics[width=\linewidth]{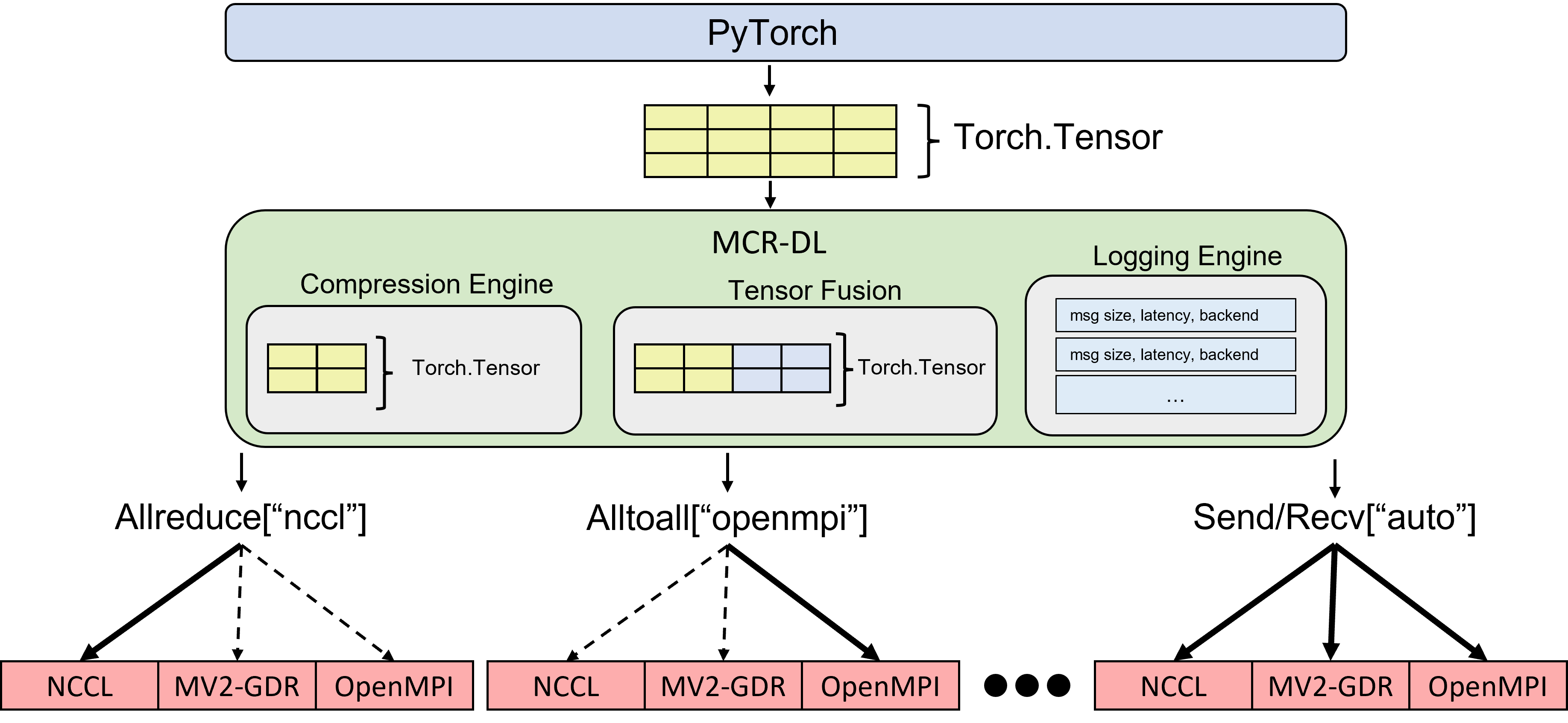}
        \caption{In MCR-DL, communication backends can be explicitly chosen, or users can dynamically choose the best backend for a given operation with the "auto" backend option. Further, MCR-DL routes all communication operations through an optional set of optimizations, including Tensor Fusion (combining small tensors into a bandwidth-optimal large tensor), message compression, and logging.} 
        \label{fig:MCR-DL}
\vspace{-3ex}
\end{figure}

\begin{figure}[thbp!]
  \begin{center}
      \mbox {
          \hspace{-1\columnsep}
          \subfigure[DS-MoE]
          {
              \includegraphics[width=.45\linewidth,trim=2 2 2 2,clip]{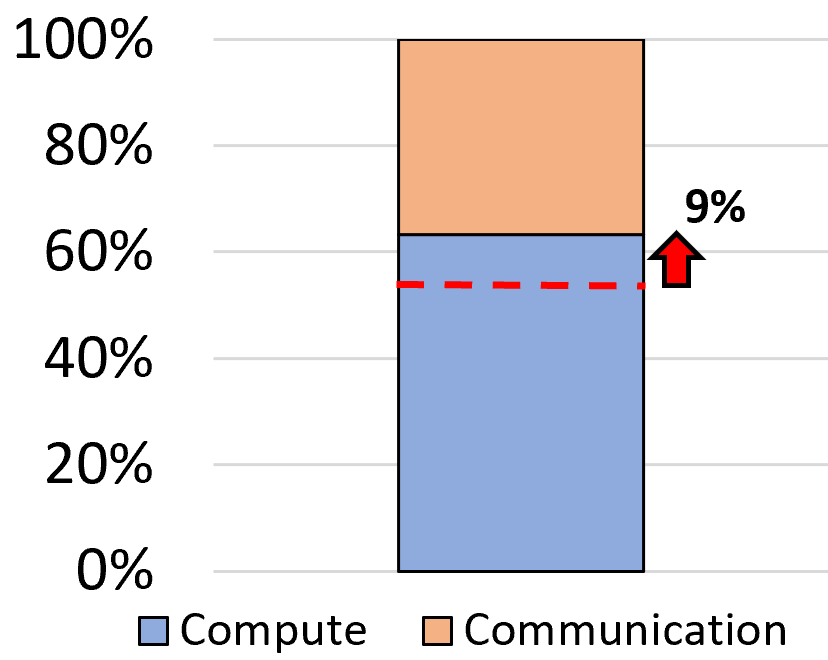}
              \label{fig:python-overhead-allreduce}
          }
          \subfigure[DLRM]
          {
              \includegraphics[width=.45\linewidth,trim=2 2 2 2,clip]{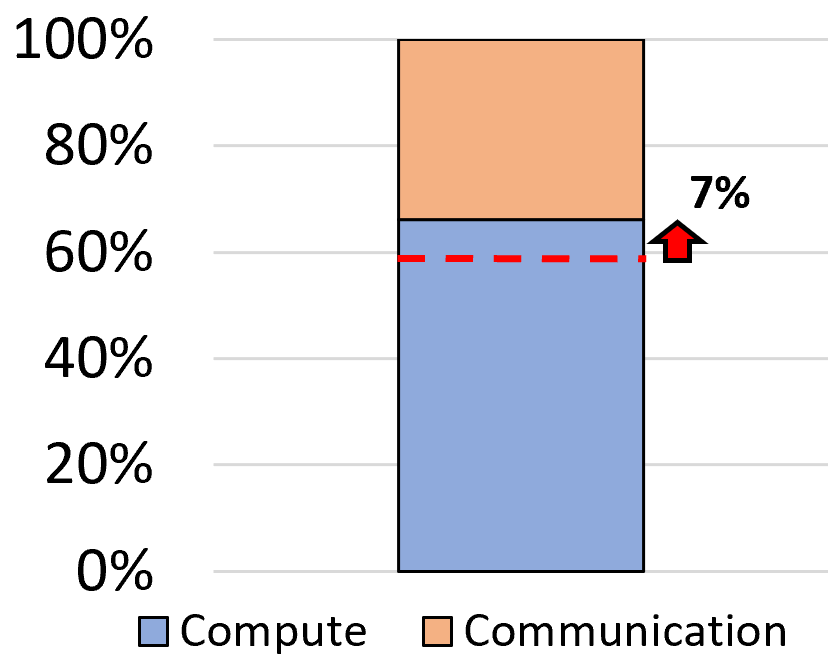}
              \label{fig:python-overhead-alltoall}
          }
      }
      \vspace*{-0.5\baselineskip}
      \caption{Communication overhead reduction with MCR-DL at 256 Lassen V100 GPUs (DS-MoE) and 32 ThetaGPU A100 GPUs (DLRM).}
      \label{fig:computevcomms-after}
  \end{center}
\vspace*{-2\baselineskip}
\end{figure}

In PyTorch's distributed module and Horovod, there are a number of commmunication optimizations (e.g. Tensor Fusion, Padding, etc) built atop the communication layer to improve performance. Similarly, by encapsulating all communication operations into MCR-DL, these optimizations can be easily integrated into all communication operations and backends. One can utilize the rich Python ecosystem to insert optimizations into MCR-DL's Python layer as depicted in Figure \ref{fig:MCR-DL}. As examples, we have implemented lossy communication compression with zfp \cite{zfp}, Tensor Fusion (combining small tensors into a bandwidth-optimal large tensor), and communication logging (which is used to generate Figures \ref{fig:comms-profiles} and \ref{fig:computevcomms-after}). Further, future optimizations (e.g. persistent collectives) can be easily added with minimal changes among backends and operations. These optimizations can be applied to incoming messages with only a few lines of Python code before routing the operation to its respective C++ backend. 

There are two parameters for Tensor Fusion: the maximum fusion buffer size $B$ and the maximum time $T$ to wait for that fusion buffer to fill with small tensors. MCR-DL introduces a small optimization for Tensor Fusion, where if the Fusion buffer does not reach $B$ before $T$ (and therefore does not saturate bandwidth), the communication is overlapped with other backends' Fusion buffers, if available. This Tensor Fusion optimization is used in all DL training results in Section \ref{sec:results}.

\subsection{Communication Tuning}
\label{sec:design-tuning}

Tuning is an established problem in distributed communication \cite{tuning-algo-selection, tuning-star, tuning-ml}. MCR-DL comes packaged with a tuning suite that seeks to map an input communication operation (with associated message size) to the best-performing backend (e.g. \textit{all\_reduce} $\rightarrow$ NCCL). This introduces additional complication since not only are distinct communication operations mixed-backend (e.g. all\_reduce and gather), but MCR-DL allows a single operation to choose the best backend with the "\textbf{auto}" backend option (e.g. mcr\_dl.gather("auto") routes \{small-message gather\} $\rightarrow$ MPI, and \{large-message gather\} $\rightarrow$ NCCL). This behavior is depicted in Figure \ref{fig:MCR-DL}, where solid lines depict the backend chosen for a given operation.

\renewcommand{\arraystretch}{1.2}
\begin{table}[htbp]
\centering
\begin{tabular}{|l|l|}
\hline
Message Size & Backend      \\ \hline
256          & MVAPICH2-GDR \\ \hline
512          & MVAPICH2-GDR \\ \hline
1024         & MVAPICH2-GDR \\ \hline
2048         & MVAPICH2-GDR \\ \hline
4096         & NCCL         \\ \hline
8192         & NCCL         \\ \hline
16384        & SCCL         \\ \hline
32768        & SCCL         \\ \hline
\end{tabular}
\vspace{1ex}
\caption{Example tuning table for the all\_gather collective operation at a single world size generated by MCR-DL}
\vspace{-1ex}
\label{tab:tuning-table}
\vspace{-3ex}
\end{table}

This tuning is implemented as a static tuning table. The tuning suite is composed of a set of micro-benchmark scripts that evaluate end-to-end time on a set of overlapped communication operations for each backend. 
By choosing the backend with the minimum end-to-end time for each input tensor size, MCR-DL generates a table like Table \ref{tab:tuning-table} for each world size (i.e. the number of GPUs) trained over.
Every collective requires its own static tuning table. 
The size of each collective's tuning table is dependent both on the number of specific message sizes we wish to tune for, as well as the number of scales (world size) we are tuning over. Specifically, a given table entry is first mapped by the world size, then by the message size.
Therefore, the total number of tuning table entries is given by:
$(\text{Num\_Collectives} \times \text{Num\_Scales} \times \text{Num\_Message\_Sizes})$. Since the performance of each communication backend depends heavily on the combination of inter-node fabric, intra-node fabric, and compute hardware used, tuning tables are not transferable across HPC systems. However, we find that general trends tend to hold across systems with a coarsely similar architecture (e.g. MVAPICH2-GDR consistently performs the best for small messages).
\section{Performance Characterization}
\label{sec:results}


\begin{figure}[!b]
\centering
    \includegraphics[width=\linewidth]{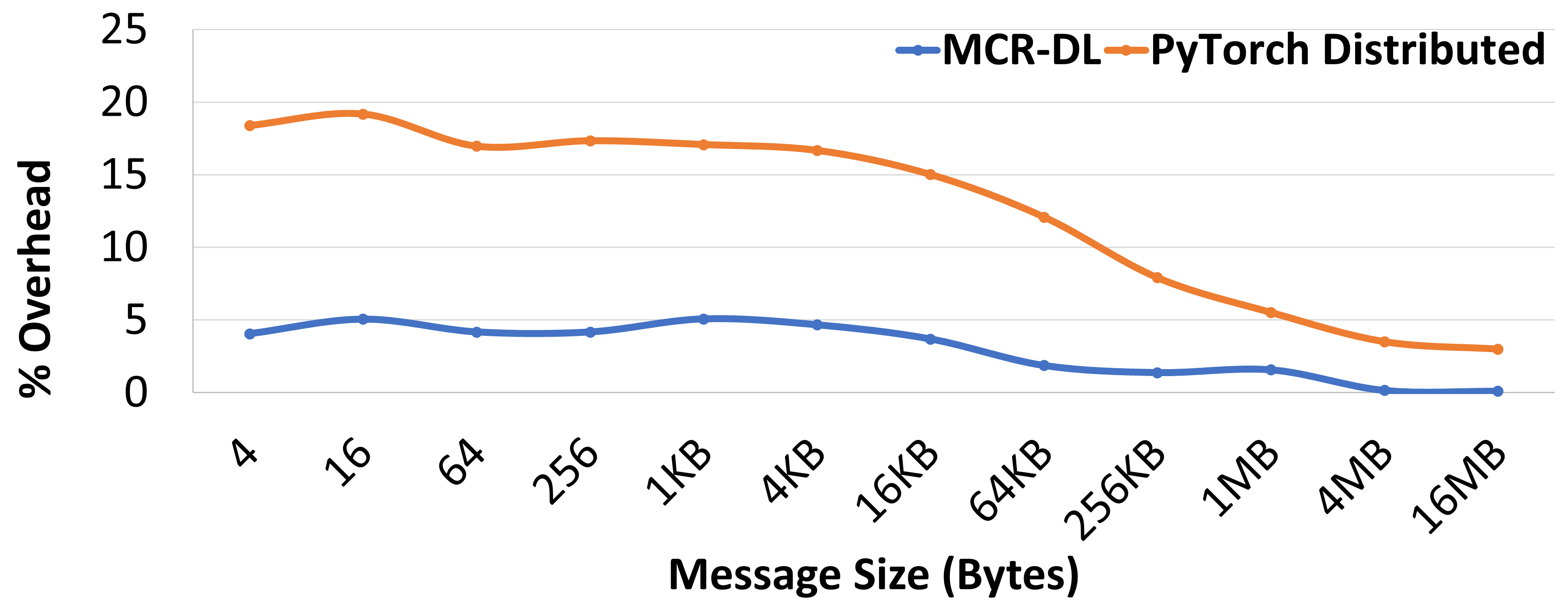}
    \caption{Overhead over OMB for MCR-DL and PyTorch Distributed for a fixed backend on ThetaGPU (32 A100 GPUs). MCR-DL reduces overhead by ensuring top-level Python logic is minimal.} 
    \label{fig:overhead}
\end{figure}

\begin{figure*}[!b]
  \begin{center}
      \mbox {
          \hspace{-1\columnsep}
          \subfigure[DS-MoE Throughput]
          {
              \includegraphics[width=.45\linewidth,trim=2 2 2 2,clip]{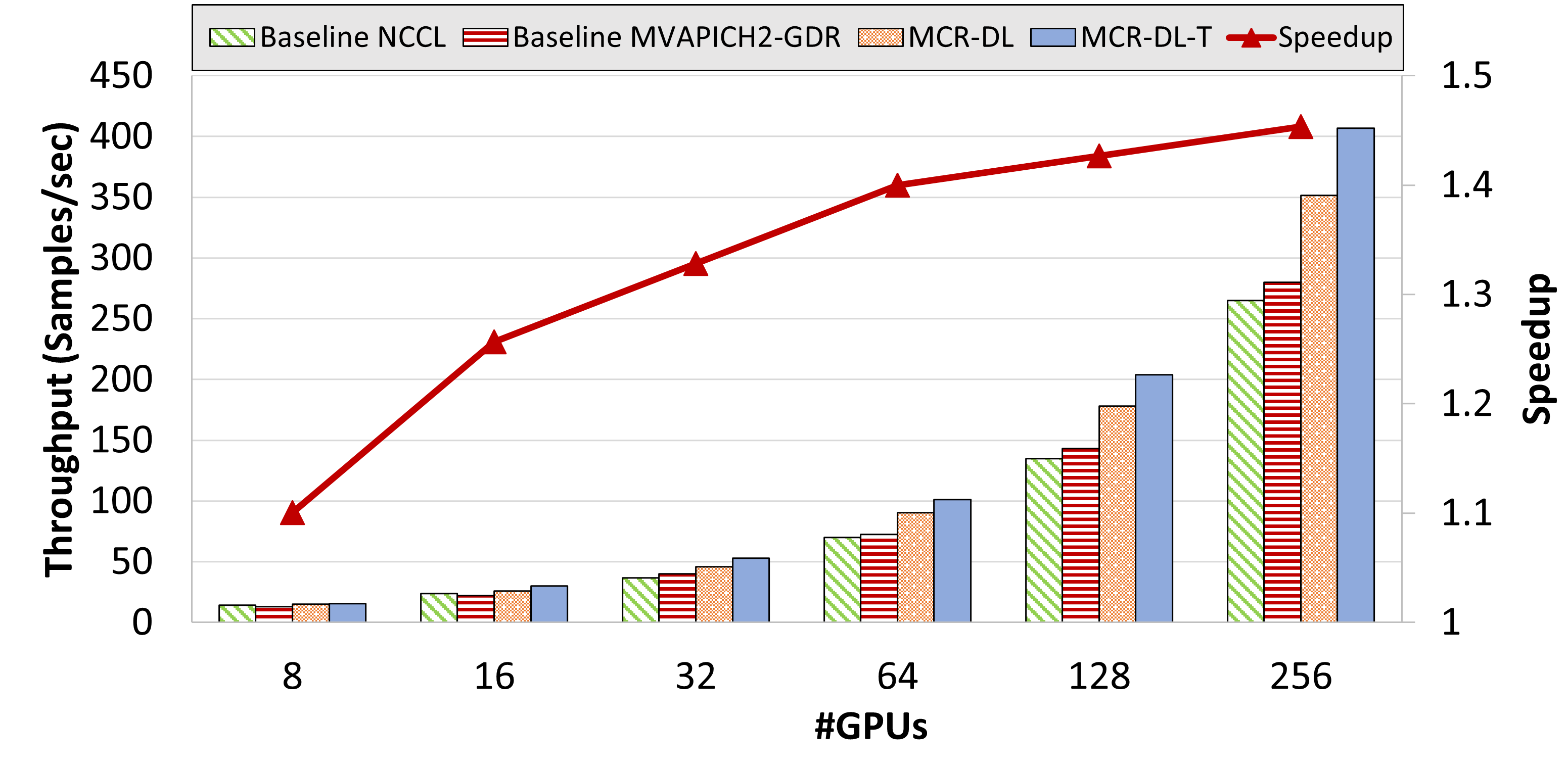}
              \label{fig:moe-throughput}
          }
          \hspace{4ex}
          \subfigure[DS-MoE Scaling Efficiency]
          {
              \includegraphics[width=.45\linewidth,trim=2 2 2 1,clip]{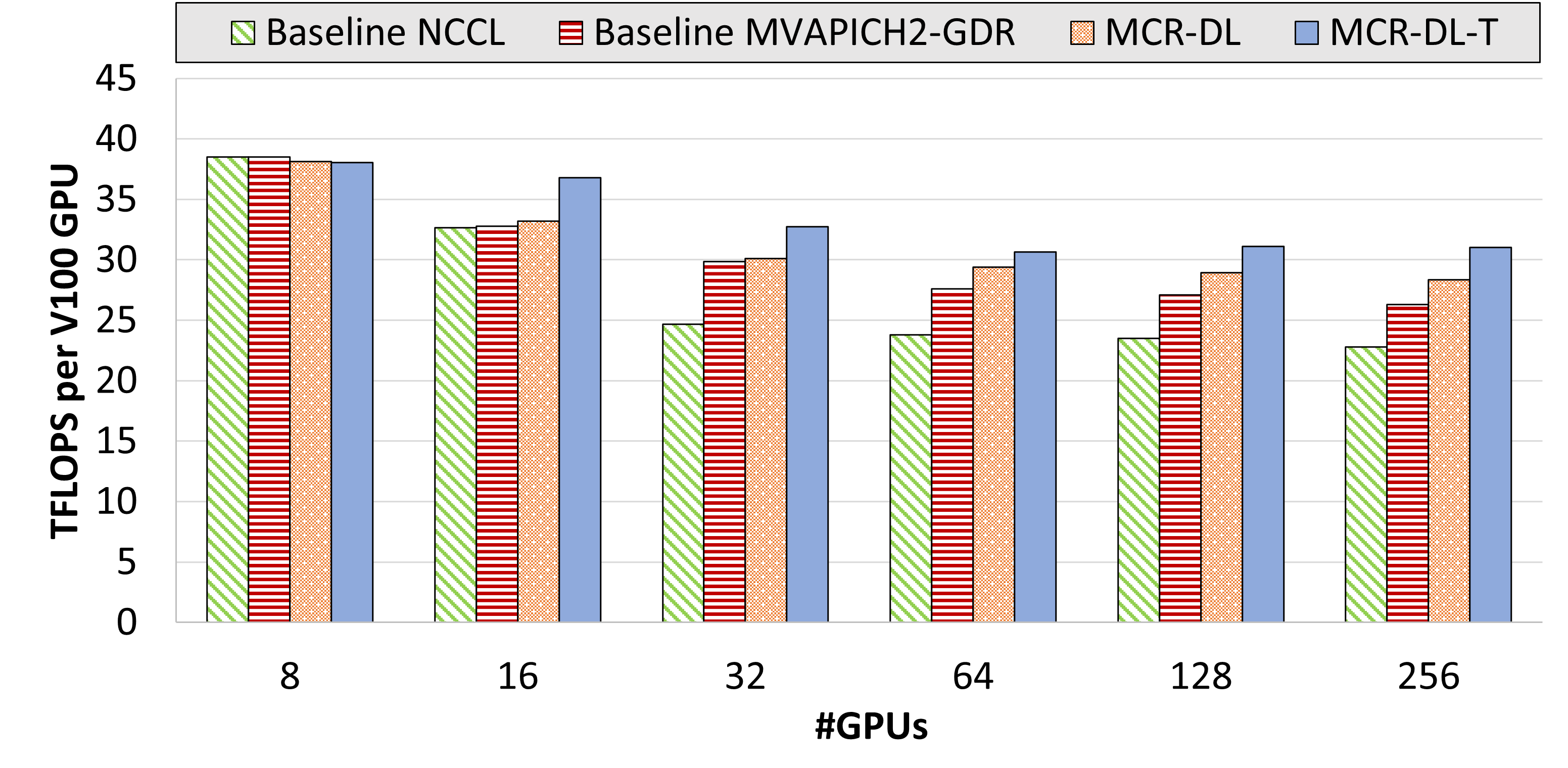}
              \label{fig:moe-efficiency}
          }
      }
      \vspace*{-0.5\baselineskip}
      \caption{Throughput and scaling efficiency improvements for DS-MoE with pure MVAPICH2-GDR, pure NCCL, and mixed-backends with MCR-DL on Lassen}
      \label{fig:moe-results}
  \end{center}
\vspace*{-1\baselineskip}
\end{figure*}

\begin{figure*}[!b]
  \begin{center}
      \mbox {
          \hspace{-1\columnsep}
          \subfigure[DLRM Throughput]
          {
              \includegraphics[width=.45\linewidth,trim=2 2 2 1,clip]{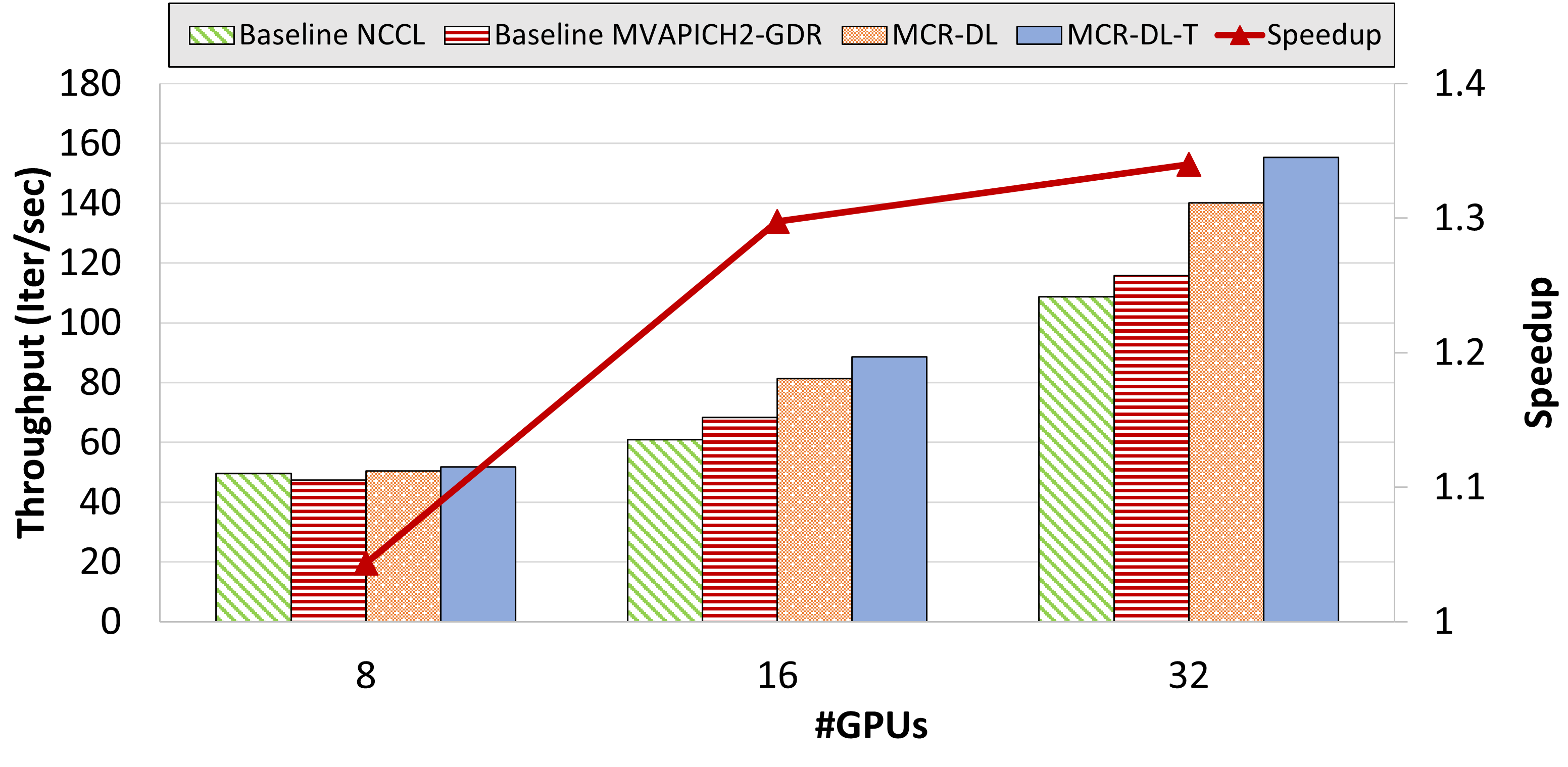}
              \label{fig:dlrm-throughput}
          }
          \hspace{4ex}
          \subfigure[DLRM Scaling Efficiency]
          {
              \includegraphics[width=.45\linewidth,trim=2 2 2 2,clip]{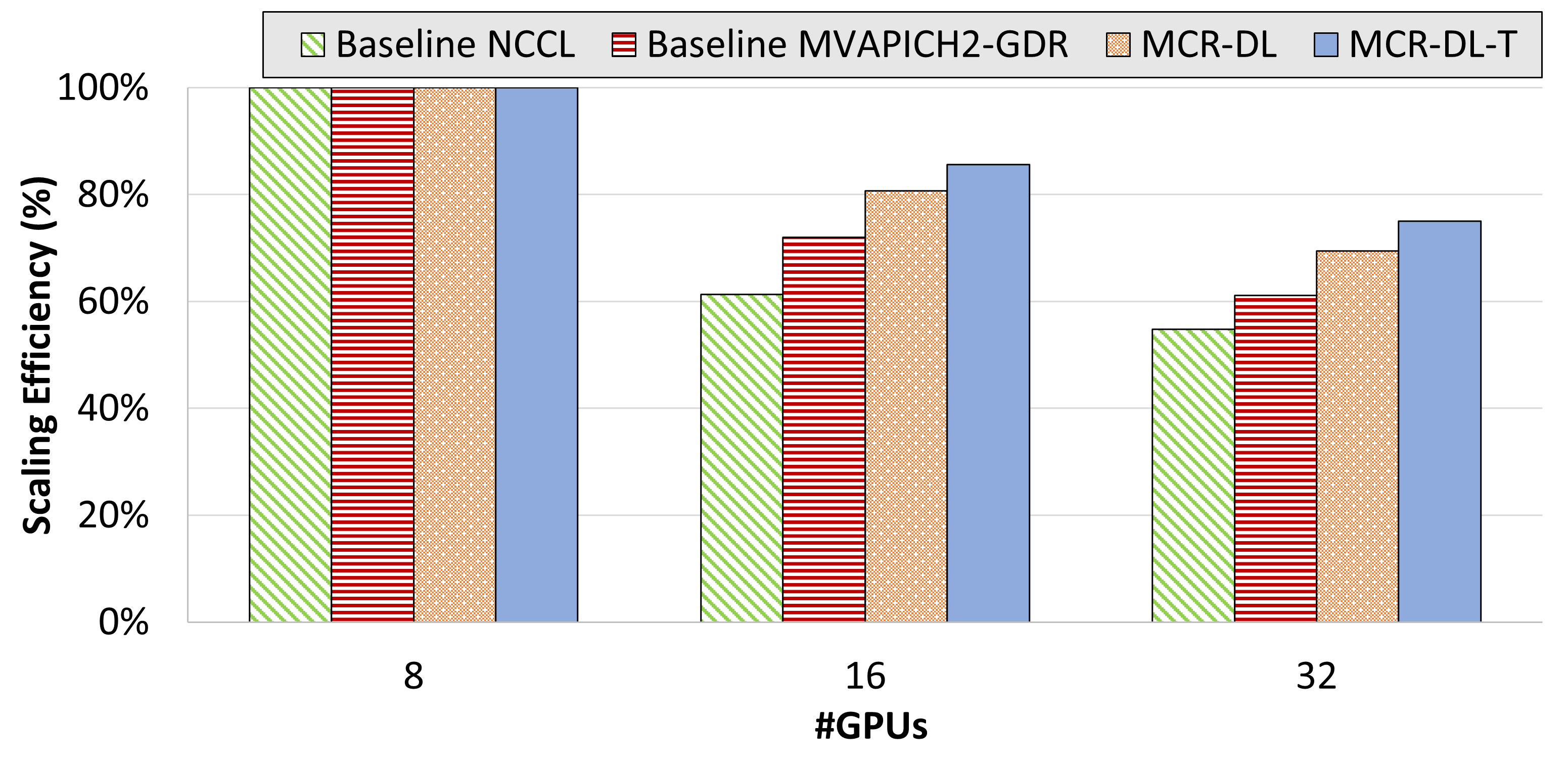}
              \label{fig:dlrm-efficiency}
          }
      }
      \vspace*{-0.5\baselineskip}
      \caption{Throughput and scaling efficiency improvements for DLRM with pure MVAPICH2-GDR, pure NCCL, and mixed-backends with MCR-DL on ThetaGPU}
      \label{fig:dlrm-results}
  \end{center}
\vspace*{-1\baselineskip}
\end{figure*}

\begin{figure*}[!b]
  \begin{center}
      \mbox {
          \hspace{-1\columnsep}
          \subfigure[Megatron-DeepSpeed Dense Model Throughput]
          {
              \includegraphics[width=.45\linewidth,trim=2 2 2 1,clip]{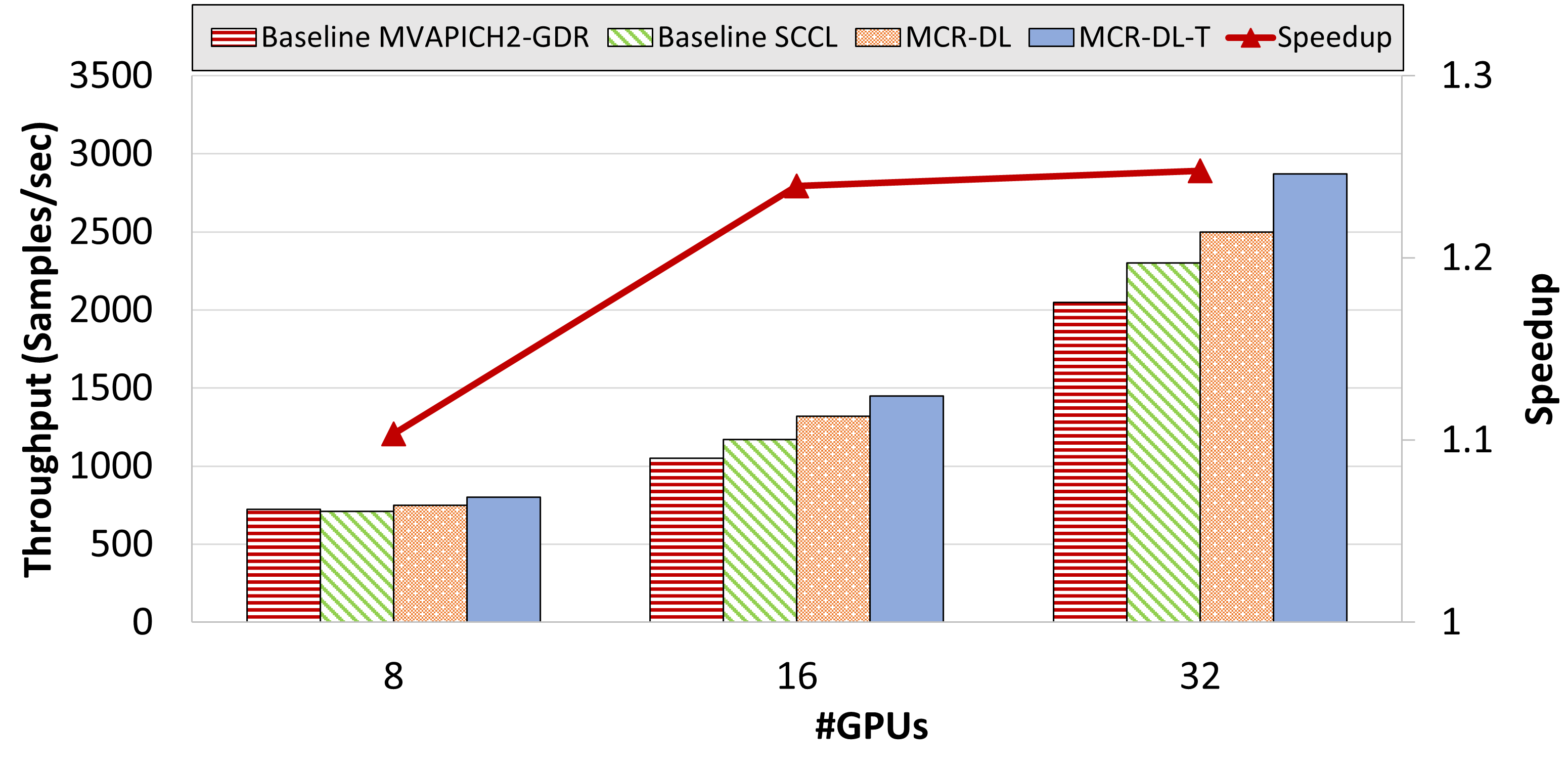}
              \label{fig:dense-throughput}
          }
          \hspace{4ex}
          \subfigure[Megatron-DeepSpeed Dense Model Scaling Efficiency]
          {
              \includegraphics[width=.45\linewidth,trim=2 2 2 1,clip]{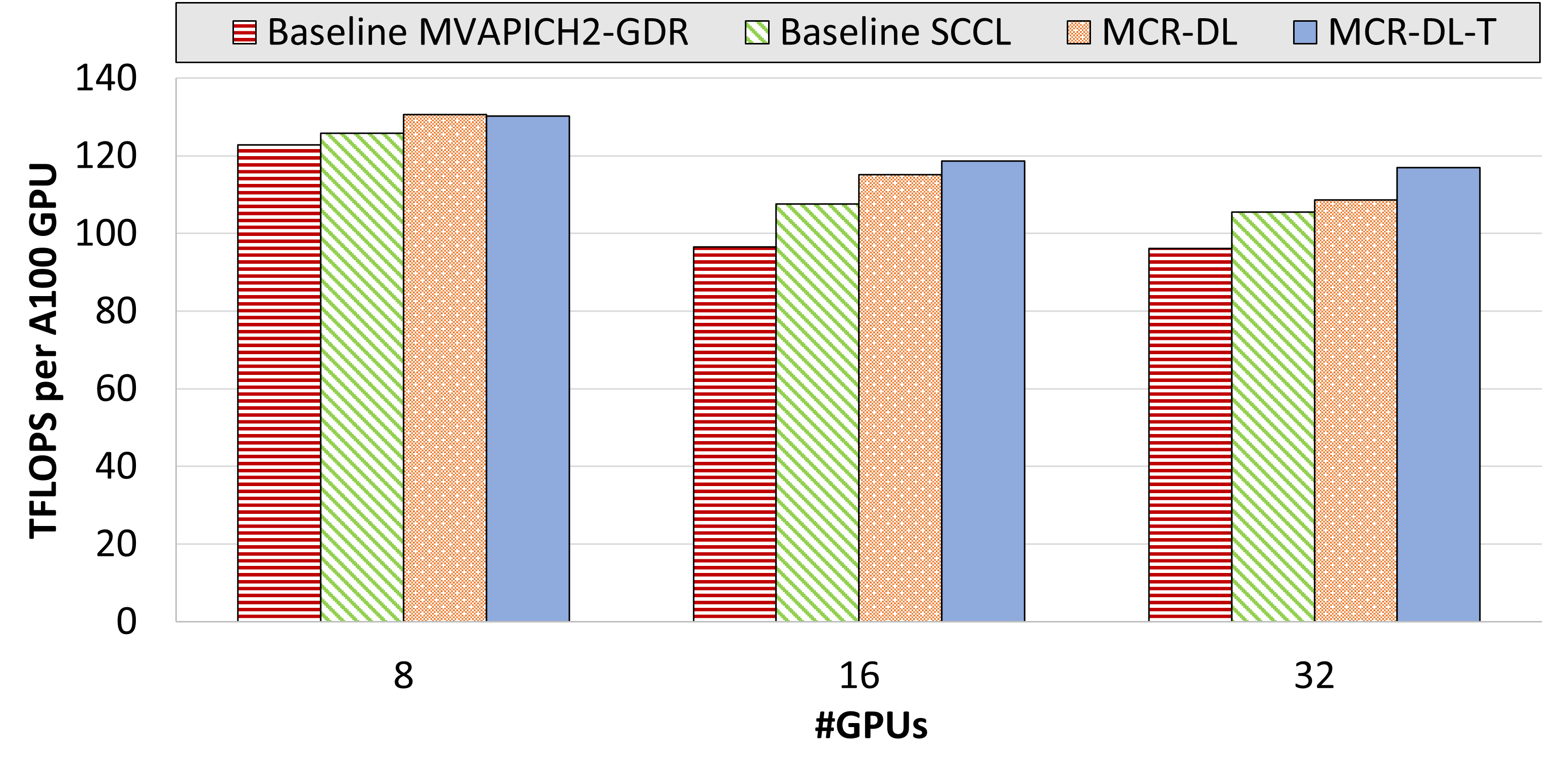}
              \label{fig:dense-efficiency}
          }
      }
      \vspace*{-0.5\baselineskip}
      \caption{Throughput and scaling efficiency improvements for dense Megatron-DeepSpeed with pure MVAPICH2-GDR, pure SCCL, and mixed-backends with MCR-DL on ThetaGPU}
      \label{fig:dense-results}
  \end{center}
\vspace*{-1\baselineskip}
\end{figure*}

\subsubsection{Node Architecture}
\label{sec:node-arch}

All experimental evaluations\footnote{The choice of cluster for a given application was purely made out of external factors such as available compute and ease of software compatibility} were carried out on the Lassen cluster at Lawrence Livermore National Laboratory and the ThetaGPU cluster at Argonne Leadership Computing Facility \cite{thetagpu}. Lassen is composed of 792 nodes each consisting of four 16 GB NVIDIA V100 GPUs and two 44-core IBM Power 9 CPUs. Nodes are connected via Mellanox Infiniband EDR in a fat-tree topology. ThetaGPU is composed of 24 NVIDIA DGX A100 nodes, each containing two AMD Rome CPUs and eight 40 GB NVIDIA A100 GPUs.

\subsubsection{Communication Backends}
\label{sec:comm-backends}

We used a mixture of MVAPICH2-GDR 2.3.7 \cite{MVAPICH2}, Open-MPI v5.1.0 \cite{openmpi} (built with UCX v1.13.1), the latest MSCCL \cite{sccl}, and NCCL 2.14.3-1 \cite{nccl} for all DL experiments. All backends and frameworks were built with CUDA 11.4.152 on ThetaGPU and CUDA 11.4.100 on Lassen.

\subsubsection{Software Libraries}
\label{sec:software}

All micro-benchmark evaluations were carried out with OSU Micro-Benchmarks (OMB) 6.1. For our DL evaluations, we used source-built PyTorch v1.12.1 and DeepSpeed v0.7.4. 

\subsubsection{DL Training Settings}
\label{sec:dl-settings}

For both DS-MoE and DLRM, we had to replace all dependencies on PyTorch's distributed module with MCR-DL calls. Since MCR-DL conforms to the PyTorch API wherever possible, this step is a straightforward search-and-replace.

We trained a 4B parameter DS-MoE model (350M+PR-MoE-32/64) on the Pile \cite{pile}. For more details on this model and on DS-MoE, see \cite{ds-moe-latest}.

For DLRM, we trained 100 synthetic data batches of size 8k with bottom and top MLPs of size (512-512-64) and (1024-1024-1024-1), respectively. The embedding table size used is 1e6 $\times$ (num\_ranks).

The dense Megatron-DeepSpeed model contained 6.7B parameters with a model-parallelism degree of 2 and ZeRO stage 2. It was also trained on the Pile \cite{pile}.

\subsection{Micro-Benchmarks}
\label{sec:results-microbench}

Before proceeding to application-level performance evaluations, we first created simple collective and point-to-point benchmarks to ensure MCR-DL doesn't introduce significant performance overhead when compared to micro-benchmarks implemented at the C-level, as investigated earlier with OMB.  As demonstrated in Figure \ref{fig:overhead}, MCR-DL introduces an overhead of around 5\% for small MPI\_Alltoall operations (under 4kB). However, this overhead quickly reduces to 1\% in the MB message range, which is the message range expected for most DL training applications \cite{hvprof}. PyTorch's distributed module built atop MVAPICH2-GDR, however, has a high overhead (18\%) for small messages, and converges to a higher overhead (4\%) in the MB message range. MCR-DL doesn't introduce significant overhead for communication operations. 

In order to spare users the OMB evaluations like Figure \ref{fig:colls-lassen-64gpu}, we created a tuning suite to generate a static tuning table for later use in applications. The tuning suite first runs basic collective and point-to-point evaluations over a range of message sizes, scales, and backends. Then, the tuning scripts create a tuning table which maps a given message size and number of processes to a given communication backend. The tables for Lassen and ThetaGPU are used in subsequent DL evaluations. This tuning table is used whenever the "auto" backend is passed to a collective as described in Section \ref{sec:design}. The difference between static-backend mixing and tuned mixing is depicted in all DL training figures as MCR-DL and MCR-DL-T, respectively.
\vspace{-0.25ex}

\subsection{DL Training}
\label{results-dl-training}

With the setup described above in \ref{sec:node-arch} through \ref{sec:dl-settings}, we carried out DL training evaluations with MCR-DL on the Lassen HPC system. Baseline experiments were carried out with PyTorch's distributed module built against a single communication backend (e.g. ``Baseline SCCL`` is PyTorch distributed built with the SCCL backend). Neither tensor fusion nor compression from Section \ref{sec:design-extensibility} were used in evaluations\footnote{While we expect performance benefits from tensor fusion and compression, we wish to isolate the effect of mixing communication backends.}. Further, to compare coarse-grained mix-and-match (i.e. one backend per collective such as NCCL \textit{Allreduce} and MPI \textit{Alltoall}) against fine-grained mix-and-match (i.e. one backend per (collective, message size) pair such as NCCL \textit{Allreduce} for 1MB messages and MPI \textit{Allreduce for 512KB messages}. These two settings of MCR-DL are depicted in Figures \ref{fig:moe-results}-\ref{fig:dense-results} as \textbf{MCR-DL} and \textbf{MCR-DL-T}, respectively. First, we run pre-training throughput experiments DS-MoE for pure NCCL, pure MVAPICH2-GDR and mixed backends. Results are depicted in \ref{fig:moe-throughput}. At smaller scales, NCCL performs better than MVAPICH2-GDR because Alltoall is not yet a dominant factor in communication time. We see a crossover threshold from Allreduce-bound to Alltoall-bound communication at around 32 GPUs, beyond which MVAPICH2-GDR's improved Alltoall starts to show benefits. The performance difference between pure NCCL and pure MVAPICH2-GDR is still small, however, because NCCL's Allreduce collective is more performant than MVAPICH2-GDR's at this message range.

MCR-DL is able to exploit MVAPICH2-GDR's improved Alltoall and NCCL's improved Allreduce to perform best at all scales without deadlocks. At 256 GPUs, we see a 31\% improvement over pure MVAPICH2-GDR and a 35\% improvement over pure NCCL. Scaling efficiency \ref{fig:moe-efficiency} is also greatly improved with MCR-DL, maintaining a 81\% efficiency at 256 V100 GPUs.

Second, we have evaluated pure NCCL, pure MVAPICH2-GDR and mixed backends on the ThetaGPU HPC system for DLRM. Results are depicted in Figure \ref{fig:dlrm-throughput}. NCCL again beats MVAPICH2-GDR at small scales due to its improved Allreduce. At higher scales, MVAPICH2-GDR again starts to perform better due to Alltoall's scaling, and MCR-DL is able to use each backend's strengths to improve performance, achieving a 25\% improvement over pure MVAPICH2-GDR and a 30\% improvement over pure NCCL. Scaling efficiency is less that of DS-MoE, but still improved by MCR-DL, maintaining a 75\% efficiency at 32 A100 GPUs. 

\vspace{-1\baselineskip}
\begin{figure}[htbp]
\centering
    \includegraphics[width=\linewidth]{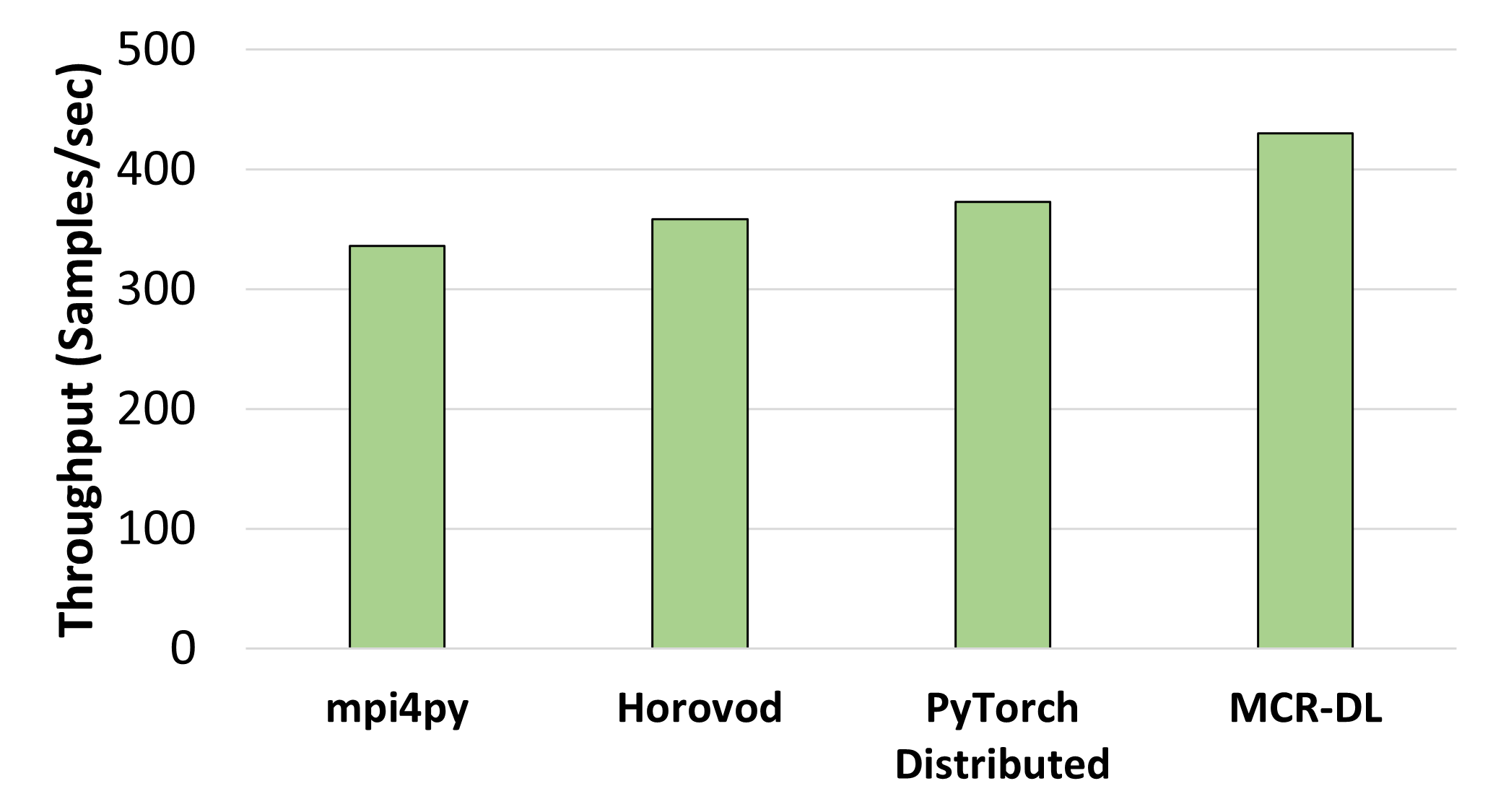}
    \vspace{-1\baselineskip}
    \caption{Comparison of MCR-DL against competing PyTorch-compatible frameworks on a Mixture-of-Experts transformer using 256 Lassen V100 GPUs.}
    \label{fig:related-results}
    \vspace{-.5\baselineskip}
\end{figure}

In order to directly compare the performance of MCR-DL with all PyTorch-compatible\footnote{LBANN does not provide any MoE implementation, and is not compatible with any mainstream DL frameworks such as PyTorch} competing frameworks in Table \ref{tab:related}, we swapped all communication operations in Megatron-DeepSpeed with each respective framework's implementation. The results on 256 Lassen V100 GPUs is depicted in Figure \ref{fig:related-results}. In order to compare each framework's best performance, MCR-DL, Horovod, and PyTorch-distributed were run with tensor fusion enabled, which leads to the performance gap between mpi4py and both Horovod and PyTorch-distributed. MCR-DL performs the best due to its mixed-backend optimizations coupled with tensor fusion.

For completeness, we have also trained a dense Megatron-DeepSpeed model on the ThetaGPU cluster with a mixture of MSCCL \cite{sccl} and MVAPICH2-GDR \cite{MVAPICH2}. As a secondary result, we have taken the compute vs. communication breakdown for DS-MoE and DLRM when using MCR-DL at 256 Lassen V100 GPUs and 32 ThetaGPU A100 GPUs, respectively. MCR-DL is an important component in reducing the computation bottleneck at scale, demonstrating a 9\% reduction in communication time for DS-MoE and a 7\% reduction in communication time for DLRM.

\section{Discussion}
\label{sec:discussion}

The throughput and scaling efficiency improvements in Figures \ref{fig:moe-results}, \ref{fig:dlrm-results}, and \ref{fig:dense-results} demonstrate that a mixed-backend DL communication framework can significantly improve the performance of emerging DL models by reducing the communication bottleneck. Further, it was confirmed that a C++ backbone underneath a thin Python layer ensures low-overhead communication operations, which enables the exploration of small-message latency-bound operations for emerging models.

These results are in agreement with the original observation that modern communication backends vary widely in performance characteristics across operations, within operations, and across releases. By mix-and-matching backends for a given operation (and within an operation), significant communication performance improvements were achieved. Further, since our communication operations are implemented in low-latency C++ code underneath a thin Python interface, we have maintained low overhead while ensuring compatibility with Python-based DL frameworks.

The performance improvements inherent in mixing communication backends are consistent with the findings of previous NCCL and MPI studies \cite{nithin-thesis} and studies exploring the mixture of MPI with external runtimes in \cite{jose-upc}. 
\section{Conclusion}
\label{sec:conclusion}

State-of-the-art deep learning (DL) models are pushing the boundaries of existing fields while pioneering entirely new areas of study. However, such DL models are often impossible or impractical to train on single processors or small-scale workstations. Further work in novel parallelism schemes and optimizations will require a robust and extensible interface between DL frameworks and communication backends. In this paper, we present and evaluate MCR-DL: a Mix-and-Match Communication Runtime for DL. MCR-DL supports all communication operations and backends, and enables mixed-backend communication to ensure the most performant backend is being used for a given communication operation. The proposed design is demonstrated on state-of-the-art DL models such as DLRM \cite{dlrm} and Mixture-of-Experts (MoE) \cite{ds-moe-orig, ds-moe-latest}. We report up to a 31\% improvement in DeepSpeed-MoE throughput on 256 V100 GPUs on the Lassen HPC system and a 25\% improvement in DLRM on 32 A100 GPUs on the Theta-GPU HPC system. We believe that MCR-DL will pave the way for designing and implementing future DL communication enhancements and distributed DL frameworks.

\bibliographystyle{IEEEtran}
\bibliography{bibfiles/bibfile}

\begin{thebibliography}{10}
\providecommand{\url}[1]{#1}
\csname url@samestyle\endcsname
\providecommand{\newblock}{\relax}
\providecommand{\bibinfo}[2]{#2}
\providecommand{\BIBentrySTDinterwordspacing}{\spaceskip=0pt\relax}
\providecommand{\BIBentryALTinterwordstretchfactor}{4}
\providecommand{\BIBentryALTinterwordspacing}{\spaceskip=\fontdimen2\font plus
\BIBentryALTinterwordstretchfactor\fontdimen3\font minus
  \fontdimen4\font\relax}
\providecommand{\BIBforeignlanguage}[2]{{%
\expandafter\ifx\csname l@#1\endcsname\relax
\typeout{** WARNING: IEEEtran.bst: No hyphenation pattern has been}%
\typeout{** loaded for the language `#1'. Using the pattern for}%
\typeout{** the default language instead.}%
\else
\language=\csname l@#1\endcsname
\fi
#2}}
\providecommand{\BIBdecl}{\relax}
\BIBdecl

\bibitem{zero-offload}
J.~Ren, S.~Rajbhandari, R.~Y. Aminabadi, O.~Ruwase, S.~Yang, M.~Zhang, D.~Li,
  and Y.~He, ``Zero-offload: Democratizing billion-scale model training,''
  2021.

\bibitem{megatron-lm}
\BIBentryALTinterwordspacing
M.~Shoeybi, M.~Patwary, R.~Puri, P.~LeGresley, J.~Casper, and B.~Catanzaro,
  ``Megatron-lm: Training multi-billion parameter language models using model
  parallelism,'' \emph{CoRR}, vol. abs/1909.08053, 2019. [Online]. Available:
  \url{http://arxiv.org/abs/1909.08053}
\BIBentrySTDinterwordspacing

\bibitem{dlrm}
M.~Naumov, D.~Mudigere, H.-J.~M. Shi, J.~Huang, N.~Sundaraman, J.~Park,
  X.~Wang, U.~Gupta, C.-J. Wu, A.~G. Azzolini \emph{et~al.}, ``Deep learning
  recommendation model for personalization and recommendation systems,''
  \emph{arXiv preprint arXiv:1906.00091}, 2019.

\bibitem{ds-moe-orig}
Y.~J. Kim, A.~A. Awan, A.~Muzio, A.~F.~C. Salinas, L.~Lu, A.~Hendy,
  S.~Rajbhandari, Y.~He, and H.~H. Awadalla, ``Scalable and efficient moe
  training for multitask multilingual models,'' \emph{arXiv preprint
  arXiv:2109.10465}, 2021.

\bibitem{ds-moe-latest}
S.~Rajbhandari, C.~Li, Z.~Yao, M.~Zhang, R.~Y. Aminabadi, A.~A. Awan,
  J.~Rasley, and Y.~He, ``Deepspeed-moe: Advancing mixture-of-experts inference
  and training to power next-generation ai scale,'' \emph{arXiv preprint
  arXiv:2201.05596}, 2022.

\bibitem{gems}
A.~Jain, A.~A. Awan, A.~M. Aljuhani, J.~M. Hashmi, Q.~G. Anthony, H.~Subramoni,
  D.~K. Panda, R.~Machiraju, and A.~Parwani, ``Gems: Gpu-enabled memory-aware
  model-parallelism system for distributed dnn training,'' in \emph{Proceedings
  of the International Conference for High Performance Computing, Networking,
  Storage and Analysis}, ser. SC '20.\hskip 1em plus 0.5em minus 0.4em\relax
  IEEE Press, 2020.

\bibitem{megatron-turing-nlg}
S.~Smith, M.~Patwary, B.~Norick, P.~LeGresley, S.~Rajbhandari, J.~Casper,
  Z.~Liu, S.~Prabhumoye, G.~Zerveas, V.~Korthikanti, E.~Zhang, R.~Child, R.~Y.
  Aminabadi, J.~Bernauer, X.~Song, M.~Shoeybi, Y.~He, M.~Houston, S.~Tiwary,
  and B.~Catanzaro, ``Using deepspeed and megatron to train megatron-turing nlg
  530b, a large-scale generative language model,'' 2022.

\bibitem{meta-opt}
\BIBentryALTinterwordspacing
S.~Zhang, S.~Roller, N.~Goyal, M.~Artetxe, M.~Chen, S.~Chen, C.~Dewan, M.~Diab,
  X.~Li, X.~V. Lin, T.~Mihaylov, M.~Ott, S.~Shleifer, K.~Shuster, D.~Simig,
  P.~S. Koura, A.~Sridhar, T.~Wang, and L.~Zettlemoyer, ``Opt: Open pre-trained
  transformer language models,'' 2022. [Online]. Available:
  \url{https://arxiv.org/abs/2205.01068}
\BIBentrySTDinterwordspacing

\bibitem{dlrm-scale}
D.~Mudigere, Y.~Hao, J.~Huang, Z.~Jia, A.~Tulloch, S.~Sridharan, X.~Liu,
  M.~Ozdal, J.~Nie, J.~Park \emph{et~al.}, ``Software-hardware co-design for
  fast and scalable training of deep learning recommendation models,''
  \emph{arXiv preprint arXiv:2104.05158}, 2021.

\bibitem{ResNet}
\BIBentryALTinterwordspacing
K.~He, X.~Zhang, S.~Ren, and J.~Sun, ``Deep residual learning for image
  recognition,'' \emph{CoRR}, vol. abs/1512.03385, 2015. [Online]. Available:
  \url{http://arxiv.org/abs/1512.03385}
\BIBentrySTDinterwordspacing

\bibitem{zero}
S.~Rajbhandari, J.~Rasley, O.~Ruwase, and Y.~He, ``Zero: Memory optimizations
  toward training trillion parameter models,'' 2020.

\bibitem{deepspeed}
\BIBentryALTinterwordspacing
J.~Rasley, S.~Rajbhandari, O.~Ruwase, and Y.~He, \emph{DeepSpeed: System
  Optimizations Enable Training Deep Learning Models with Over 100 Billion
  Parameters}.\hskip 1em plus 0.5em minus 0.4em\relax New York, NY, USA:
  Association for Computing Machinery, 2020, p. 3505–3506. [Online].
  Available: \url{https://doi.org/10.1145/3394486.3406703}
\BIBentrySTDinterwordspacing

\bibitem{gshard}
D.~Lepikhin, H.~Lee, Y.~Xu, D.~Chen, O.~Firat, Y.~Huang, M.~Krikun, N.~Shazeer,
  and Z.~Chen, ``Gshard: Scaling giant models with conditional computation and
  automatic sharding,'' 2020.

\bibitem{nccl}
\BIBentryALTinterwordspacing
{NVIDIA}, ``{NVIDIA Collective Communications Library (NCCL)},'' 2016,
  {Accessed: \today}. [Online]. Available:
  \url{https://developer.nvidia.com/nccl}
\BIBentrySTDinterwordspacing

\bibitem{openmpi}
{The Open MPI Development Team}, ``{{Open MPI : Open Source High Performance
  Computing}},'' http://www.open-mpi.org, 2004, [Online; accessed \today].

\bibitem{MVAPICH2}
{MVAPICH2: MPI over InfiniBand, 10GigE/iWARP and RoCE},
  https://mvapich.cse.ohio-state.edu/, 2001, [Online; accessed \today].

\bibitem{mpi4py}
L.~Dalcin and Y.-L.~L. Fang, ``mpi4py: Status update after 12 years of
  development,'' \emph{Computing in Science Engineering}, vol.~23, no.~4, pp.
  47--54, 2021.

\bibitem{pytorch-dist}
\BIBentryALTinterwordspacing
S.~Li, Y.~Zhao, R.~Varma, O.~Salpekar, P.~Noordhuis, T.~Li, A.~Paszke,
  J.~Smith, B.~Vaughan, P.~Damania, and S.~Chintala, ``Pytorch distributed:
  Experiences on accelerating data parallel training,'' \emph{CoRR}, vol.
  abs/2006.15704, 2020. [Online]. Available:
  \url{https://arxiv.org/abs/2006.15704}
\BIBentrySTDinterwordspacing

\bibitem{horovod}
\BIBentryALTinterwordspacing
A.~Sergeev and M.~Del~Balso, ``{Horovod: Fast and Easy Distributed Deep
  Learning in TensorFlow},'' \emph{CoRR}, vol. abs/1802.05799, 2018. [Online].
  Available: \url{http://arxiv.org/abs/1802.05799}
\BIBentrySTDinterwordspacing

\bibitem{jose-upc}
\BIBentryALTinterwordspacing
J.~Jose, M.~Luo, S.~Sur, and D.~K. Panda, ``Unifying upc and mpi runtimes:
  Experience with mvapich,'' in \emph{Proceedings of the Fourth Conference on
  Partitioned Global Address Space Programming Model}, ser. PGAS '10.\hskip 1em
  plus 0.5em minus 0.4em\relax New York, NY, USA: Association for Computing
  Machinery, 2010. [Online]. Available:
  \url{https://doi.org/10.1145/2020373.2020378}
\BIBentrySTDinterwordspacing

\bibitem{aluminum}
N.~Dryden, N.~Maruyama, T.~Moon, T.~Benson, A.~Yoo, M.~Snir, and B.~Van~Essen,
  ``Aluminum: An asynchronous, gpu-aware communication library optimized for
  large-scale training of deep neural networks on hpc systems,'' in \emph{2018
  IEEE/ACM Machine Learning in HPC Environments (MLHPC)}, 2018, pp. 1--13.

\bibitem{vit-moe}
\BIBentryALTinterwordspacing
C.~Riquelme, J.~Puigcerver, B.~Mustafa, M.~Neumann, R.~Jenatton, A.~S. Pinto,
  D.~Keysers, and N.~Houlsby, ``Scaling vision with sparse mixture of
  experts,'' \emph{CoRR}, vol. abs/2106.05974, 2021. [Online]. Available:
  \url{https://arxiv.org/abs/2106.05974}
\BIBentrySTDinterwordspacing

\bibitem{mtf}
N.~Shazeer, Y.~Cheng, N.~Parmar, D.~Tran, A.~V. \textit{et al}
  B.~A. Hechtman.

\bibitem{pytorch}
A.~Paszke, S.~Gross, S.~Chintala, G.~Chanan, E.~Yang, Z.~DeVito, Z.~Lin,
  A.~Desmaison, L.~Antiga, and A.~Lerer, ``{Automatic Differentiation in
  PyTorch},'' 2017.

\bibitem{spectrum-mpi}
{IBM}, ``{IBM Spectrum MPI: Accelerating high-performance application
  parallelization},'' https://www.ibm.com/us-en/marketplace/spectrum-mpi, 2018,
  {Accessed: \today}.

\bibitem{Kawthar:IWOPH19}
K.~S. Khorassani, C.-H. Chu, H.~Subramoni, and D.~K. Panda, ``{Performance
  Evaluation of MPI Libraries on GPU-enabled OpenPOWER Architectures: Early
  Experiences},'' in \emph{International Workshop on OpenPOWER for HPC (IWOPH
  19) at the 2019 ISC High Performance Conference}, 2018.

\bibitem{sccl}
\BIBentryALTinterwordspacing
Z.~Cai, Z.~Liu, S.~Maleki, M.~Musuvathi, T.~Mytkowicz, J.~Nelson, and
  O.~Saarikivi, ``Synthesizing optimal collective algorithms,'' \emph{CoRR},
  vol. abs/2008.08708, 2020. [Online]. Available:
  \url{https://arxiv.org/abs/2008.08708}
\BIBentrySTDinterwordspacing

\bibitem{fairseq-moe}
M.~Artetxe, S.~Bhosale, N.~Goyal, T.~Mihaylov, M.~Ott, S.~Shleifer, X.~V.~L.
  \textit{et al} 
  S.~Chen, H.~Akin, M.~Baines, L.~Martin, X.~Zhou, P.~S. Koura, B.~O'Horo,
  J.~Wang, L.~Zettlemoyer, M.~T. Diab, Z.~Kozareva, and V.~Stoyanov.

\bibitem{stream-mpich}
\BIBentryALTinterwordspacing
H.~Zhou, K.~Raffenetti, Y.~Guo, and R.~Thakur, ``{MPIX} stream: An explicit
  solution to hybrid mpi programming,'' in
  \emph{{EuroMPI}/{USA}{\textquotesingle}22: 29th European {MPI}
  Users{\textquotesingle} Group Meeting}.\hskip 1em plus 0.5em minus
  0.4em\relax {ACM}, sep 2022. [Online]. Available:
  \url{https://doi.org/10.1145\%2F3555819.3555820}
\BIBentrySTDinterwordspacing

\bibitem{zfp}
P.~Lindstrom, ``Fixed-rate compressed floating-point arrays,'' \emph{IEEE
  Transactions on Visualization and Computer Graphics}, vol.~20, no.~12, pp.
  2674--2683, 2014.

\bibitem{tuning-algo-selection}
J.~Pje{\v{s}}ivac-Grbovi{\'c}, G.~Bosilca, G.~E. Fagg, T.~Angskun, and J.~J.
  Dongarra, ``Mpi collective algorithm selection and quadtree encoding,''
  \emph{Parallel Computing}, vol.~33, no.~9, pp. 613--623, 2007.

\bibitem{tuning-star}
A.~Faraj, X.~Yuan, and D.~Lowenthal, ``Star-mpi: self tuned adaptive routines
  for mpi collective operations,'' in \emph{Proceedings of the 20th annual
  international conference on Supercomputing}, 2006, pp. 199--208.

\bibitem{tuning-ml}
S.~Hunold, A.~Bhatele, G.~Bosilca, and P.~Knees, ``Predicting mpi collective
  communication performance using machine learning,'' in \emph{2020 IEEE
  International Conference on Cluster Computing (CLUSTER)}.\hskip 1em plus
  0.5em minus 0.4em\relax IEEE, 2020, pp. 259--269.

\bibitem{thetagpu}
{Argonne National Laboratory}, ``{Theta/ThetaGPU Machine Overview},''
  https://www.alcf.anl.gov/support-center/theta/theta-thetagpu-overview, 2021,
  {Accessed: \today}.

\bibitem{pile}
\BIBentryALTinterwordspacing
L.~Gao, S.~Biderman, S.~Black, L.~Golding, T.~Hoppe, C.~Foster, J.~Phang,
  H.~He, A.~Thite, N.~Nabeshima, S.~Presser, and C.~Leahy, ``The pile: An 800gb
  dataset of diverse text for language modeling,'' \emph{CoRR}, vol.
  abs/2101.00027, 2021. [Online]. Available:
  \url{https://arxiv.org/abs/2101.00027}
\BIBentrySTDinterwordspacing

\bibitem{hvprof}
A.~A. Awan, A.~Jain, C.-H. Chu, H.~Subramoni, and D.~Panda, ``{Communication
  Profiling and Characterization of Deep Learning Workloads on Clusters with
  High-Performance Interconnects},'' in \emph{Hot Interconnects 26 (HotI '19)},
  August 2019.

\bibitem{nithin-thesis}
\BIBentryALTinterwordspacing
N.~Senthil~Kumar, ``Designing optimized mpi+nccl hybrid collective
  communication routines for dense many-gpu clusters,'' Master's thesis, 2021.
  [Online]. Available:
  \url{http://rave.ohiolink.edu/etdc/view?acc\_num=osu1619132252608831}
\BIBentrySTDinterwordspacing

\end{thebibliography}

\end{document}